\documentclass[12pt]{spieman}  

\usepackage[numbers]{natbib}
\usepackage{amsmath,amsfonts,amssymb}
\usepackage{graphicx}
\usepackage{setspace}
\usepackage{tocloft}
\usepackage{amsmath,amsfonts,amssymb}
\usepackage{graphicx}
\usepackage{setspace}
\usepackage{tocloft}
\usepackage{cite}
\usepackage{amsmath,amssymb,amsfonts}
\usepackage{textcomp}
\usepackage{amsfonts,amssymb}
\usepackage{setspace}
\usepackage{listings}
\usepackage{amsmath}
\usepackage[export]{adjustbox}
\usepackage{algorithm, algpseudocode}
\usepackage{caption,setspace}
\usepackage{comment}
\usepackage{subcaption}
\usepackage{wrapfig}
\usepackage{multicol}
\usepackage{multirow}
\usepackage{epstopdf}
\renewcommand{\deg}{^{\circ}}
\newcommand{\as}{^{\prime\prime}}
\newcommand{\am}{^{\prime}}
\newcommand{\etal}{\textit{et al.\,}\,}

\cftpagenumbersoff{figure}
\cftpagenumbersoff{table}

\title{Low-Cost Raspberry Pi Star Sensor for Small Satellites}

\author[a,b,*]{Bharat Chandra P}
\author[c]{Mayuresh Sarpotdar}
\author[a]{Binukumar G. Nair}
\author[a]{Richa Rai}
\author[a]{Rekhesh Mohan}
\author[d]{Joice Mathew}
\author[a]{Margarita Safonova}
\author[a]{Jayant Murthy}
\affil[a]{Indian Institute of Astrophysics, Bangalore, 560034, India.}
\affil[b]{University of Calcutta, Kolkata, 700073, India. }
\affil[c]{Satellite Research Centre, Nanyang Technological University, Singapore.}
\affil[d]{Advanced Instrumentation and Technology Centre, RSAA, Mount Stromlo Observatory, Australian National University, Canberra, Australia.}

\cftpagenumbersoff{figure}
\cftpagenumbersoff{table}

\begin{document}

\maketitle

\begin{abstract}
We present here a low-cost Raspberry Pi (RPi)-based star sensor {\it StarberrySense} using commercial-off-the-shelf (COTS) components, developed and built for applications in small satellites and CubeSat-based missions. 
A star sensor is one of the essential instruments onboard a satellite for attitude determination. 
However, most commercially available star sensors are expensive and bulky to be used in small satellite missions. 
{\it StarberrySense} is a configurable system -- it can operate as an imaging camera, a centroiding camera, or as a star sensor. We describe the algorithms implemented in the sensor, its assembly and calibration. This payload was selected by a recent Announcement of Opportunity call for payloads to fly on the PS4-Orbital Platform by the Indian Space Research Organization (ISRO).
\end{abstract}

\keywords{Star Sensor, Attitude Sensor, Small Satellites, Raspberry Pi, CubeSat}

{\noindent \footnotesize\textbf{*}Bharat Chandra P,  \linkable{bharat.chandra@iiap.res.in} }

\begin{spacing}{1}   

\section{Introduction}
\label{sect:intro}
A star sensor onboard a satellite determines the satellite's attitude in space by identifying the stars in its field of view (FOV). The advantage of a star sensor over other sensors such as magnetometers, sun sensors, etc., is that it provides more precise orientation information of the satellite \cite{star}, to order of a few arcseconds to micro-arcseconds depending on the star sensor model. In recent years, small satellites are marking a new era of space exploration. Among these, a class of small satellites is called the CubeSats. They are usually used in low-Earth orbits (LEO) for communication, remote sensing, scientific missions and more. There might be a need for pointing information in these kinds of missions to perform orientation corrections to perform a particular observation.
Most of the commercial radiation-hardened star sensors are made for long-duration space missions. However, using these expensive star sensors is not economical for the low-budget short-duration CubeSat or any SmallSat mission in LEO, where radiation levels are low\cite{rad_level}. Also, in some cases, commercial star sensors are bulky and consume a lot of power, making them impractical to be used in CubeSats\cite{Lee20}. To address this problem, our group started the development of a low-cost star sensor \cite{mico} using either readily available components or even building some of them in-house. The first prototype, {\it StarSense}, was built using a Star 1000 CMOS detector, a MIL-grade FPGA board, and a custom-fabricated lens system at the cost of around 10,000 USD \cite{mico_cam}. Sarpotdar \etal had also developed a software package to evaluate the performance of star sensor algorithms, which included parameters such as attitude accuracy, sky coverage, catalogue size, etc., for different values of focal length, field of view, distortion effects and other attributes. We have continued the development of low-cost star sensors for short duration missions using readily available off-the-shelf components with a fast development cycle, modular design, and easy source code portability to different hardware. One concept of such a low-cost star sensor development, called Lab Open Star Tracker (SOST), based on Raspberry Pi and using existing open-source astronomy software, was recently described by Gutierrez \etal \cite{sost}. They employed the \textit{Source Extractor} to extract the bright sources from the captured image and \textit{Match} for matching the image of the sky with the projected segments of the sky stored onboard, followed by the attitude determination. From the real-sky test, they estimated the sensor accuracy to be about $30\as - 60\as$ with an update rate of 0.05 Hz. 

In {\it StarberrySense} we have used Raspberry Pi(RPi) Zero \cite{rasp} as the main controller costing around 15 dollars, along with an RPi camera as the detector, off-the-shelf lens system for imaging, a custom-made power supply and a custom-made housing to mount all the component, bringing the total price down to about 1600 USD. Raspberry Pi is a small single-board computer (SBC) developed by the Raspberry Pi foundation\footnote{http://www.raspberrypi.org}. It provides a compact, low-power, flexible platform to which different devices can be interfaced for applications ranging from home automation to aeronautical communication. RPi hardware design and computational capability allow it to be used in various small satellite missions, and this was the main reason for choosing the RPi. Using off-the-shelf components makes the development of a star sensor fast, cheap and easy without relying on complicated hardware and optics design. The {\it StarberrySense} has an accuracy of $30 \as$  with an update rate of 0.2 Hz, and an average power consumption under 1.25 W (see Section 2 and Table~\ref{tab:specs}). Table~\ref{comp} shows the comparison between the {\it StarberrySense} and other commercially available star sensors\cite{st-16,comptech,PROCYON&RIGEL}.\\
The software for the {\it StarberrySense} is written in standard C/C++ for faster processing on a free, open-source (FOSS) Linux platform and is easily portable to any other system such as Embedded Linux or RTOS (Real-Time Operating System), allowing faster development and deployment. The modular code architecture allows for easy customization, which is vital in the design of such subsystems in CubeSat or small satellite missions. The sensor operates in Lost-In-Space (LIS) mode \cite{gv}, where centroids of possible stars in the image are identified by a region-growing algorithm. The geometric voting algorithm \cite{gv} is used to match the centroids with stars stored in the onboard catalogue. The final attitude determination is performed using the quaternion-estimator (QUEST) \cite{quest} algorithm. A minimum of three stars is required for attitude determination. The stored onboard catalogue is the Hipparchus bright star catalogue \cite{hip}, listing stars with magnitudes up to 6.5.\\
 
\begin{center}
    
\begin{table}[]

\footnotesize
\centering
\begin{tabular}{l l l l l l l}
\hline
Star Sensor                    & Accuracy(deg)  & Update Rate(Hz)   & Cost(\$)  & Size (mm)       & Weight(kg)         &Power(W)\\ \hline
Procyon Star Tracker           &0.008          &$1-4$             &330k      &$155\times210\times56$  &1.2    &6.5  \\ 
Comtech MST                    &0.02           &1                 &150k      &$50\times80\times80$    &0.3    &2 \\ 
Sinclair Interplanetary\\ ST-16   &0.002          &2              &120k      &$59\times56\times 31.5$  &0.09   &1   \\ 
Rigel-L                        &0.006          &$5-16$            &430k      &NA                          &2.2    &10 \\ 
{\it StarberrySense}           &0.008          &0.2               &1600      &$98\times75\times69$    &0.3    &1.25  \\ 
\hline   
\end{tabular}
\caption{Comparison between {\it StarberrySense} and other star sensors available in the market.}
\label{comp}
\end{table}
\end{center}
\section{The Instrument}

\begin{figure}[H]
\centering
\begin{minipage}{.49\textwidth}
    \centering
    \includegraphics[width =0.97\linewidth]{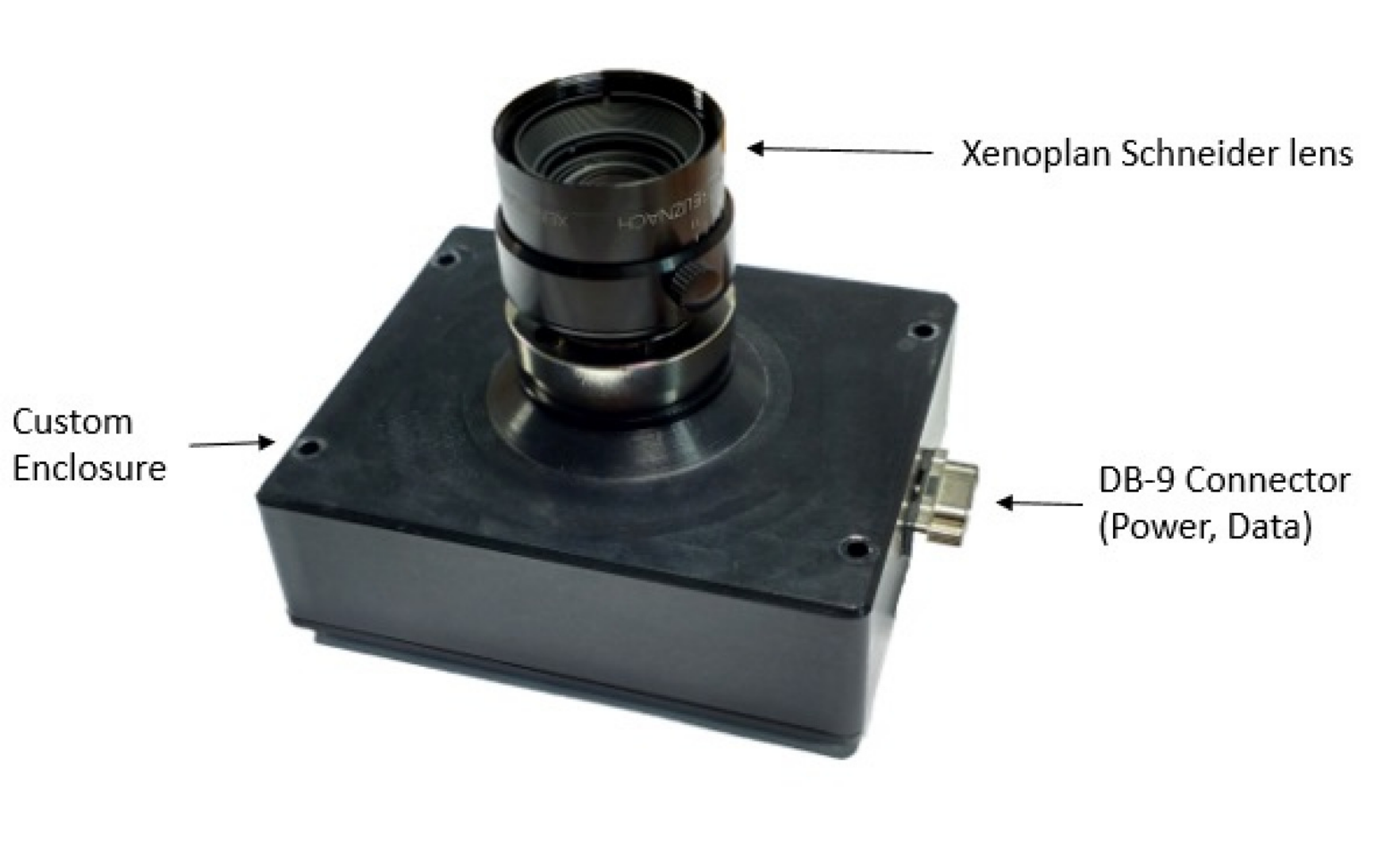}
    \caption{Photo of the star sensor without the baffle.}
    \label{fig:model}
\end{minipage}\hfill%
\begin{minipage}{.49\textwidth}
    \centering
    \includegraphics[width =\linewidth]{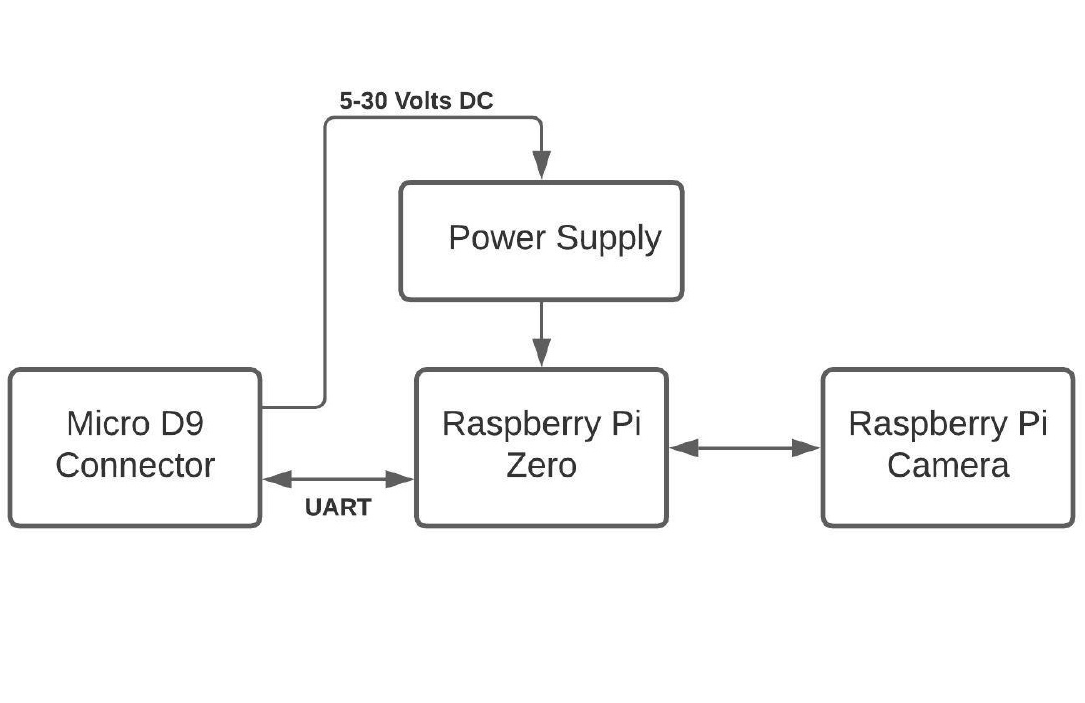}
    \caption{The electronics block diagram of the star sensor.}
    \label{fig:model-1}
\end{minipage}
\end{figure}

A star sensor is essentially an imaging camera that images the stars, processes the images, identifies the bright stars and matches their positions with the stored star catalogue for determining the look direction. As such, it consists of several subsystems that include optics, the image sensor, electronics and the housing (Fig.~\ref{fig:model}). The main controller in our star sensor is the Raspberry Pi Zero (Table~\ref{tab:rpi}).  

\begin{table}[h!]
\footnotesize

\centering
\begin{tabular}{ll}
\hline
Chip Broadcom    & BCM2835                             \\
CPU              & ARM1176JZF-S Single Core            \\                  
                 &32-bit CPU                           \\
Operating system & Raspbian                            \\
GPU              & Broadcom VideoCore IV               \\
CPU Clock        & 1 GHz                               \\
Memory           & 512 MB  DDR2                        \\
Interfaces       & $1 \times$Micro USB                 \\
                 & $1 \times$UART                      \\
I/O              & 40 GPIO Pins                        \\
                 &  CSI camera connector               \\
Onboard storage  & SD, MMC, SDIO card slot             \\
Weight           & 45 gm                                \\
Power rating     & 250 mA (1.25 Watt)                     \\
Dimensions      & 66.0mm $\times$ 30.5mm $\times$ 5.0mm   \\
\hline
\end{tabular}
\caption{Raspberry Pi Zero technical specifications.}
\label{tab:rpi}
\end{table}

\subsection{Design Requirements}

The design requirement was to build a low-cost star sensor for small satellites and CubeSat class missions. The pointing accuracy requirement was based on the sub-arcminute pointing requirement for small satellite astronomy missions. The total mass was constrained to under 500 grams, and the system's power consumption below 2 W as per the mass and power budget of CubeSats and small satellites. Since we had a stringent need for a shorter development cycle, we chose to design the star sensor from readily available off-the-shelf electronics and ruggedized optical components.
\\

For the star sensor to work, three or more stars are required to be visible in its FOV. Therefore, the FOV (which depends on the focal length of the optics and the size of the image sensor) and the magnitude detection limit must be chosen carefully to meet the requirement. For our design, the RPi V2 camera module was used as the image sensor and the Schindler Xenoplan 23-mm lens system as the imaging optics. This setup provided the system with a FOV of $9.31^{\circ}\times 7^{\circ}$, and the magnitude of the star detection limit was set to 6.5 ensuring a minimum of 3 stars in the FOV\cite{mico}.

\subsection{Image Sensor and Lens System}

The star sensor uses the Xenoplan Schneider C-mount lens system designed to work in visible and near-IR range ($400-1000$ nm). The RPi V2 No-IR camera used in the star sensor has a Sony IMX219 CMOS image sensor with no IR filter that is sensitive from visible band to near-IR band. With the 23-mm lens system and the 1.12-micron pixel sensor, the star sensor has a pixel scale of $20.5^{\as}$ per pixel (with $2\times2$ binning). The dark noise characterisation for RPi V2 camera was done by Pagnutti \etal \cite{rpi_cam}. To ensure that stars with a magnitude of 6.5 are imaged with a good SNR, we chose the exposure to be 500 ms. In this setup, the mean value of background was 1.87 counts/pixel ($\sigma=0.89$), and 6.5 mag stars were detected with $SNR=30$.

\subsection{Electronics}

The Raspberry Pi Zero with ARM1176JZF-S Single Core processor is the main controller of our star sensor, which performs the image capture, image processing, star identification and attitude determination. The Raspberry Pi is interfaced with the camera module with the MIPI CSI (Camera Serial Interface), and the Broadcom VideoCore IV, onboard RPi, handles the image readout from the Sony IMX219 CMOS sensor. The universal asynchronous receiver-transmitter (UART) port of the RPi is configured for transmitting the output data (quaternions and centroids). The RPi Zero and the camera module consume less than 1.25 watts, which is within our power limit requirement. The power supply for the star sensor was designed based on a switching mode regulator LMR33630 from Texas Instruments, which provides a 5 V regulated voltage over a wide input range from 5 to 30 V. The LMR33630 has a wide operating temperature range from -40 to +125 $ ^\circ C $, making it suitable for our application (the summary of the technical specifications is presented in Table~\ref{tab:specs}). Fig.~\ref{fig:model-1} shows the overall electronics block diagram of the star sensor.

\begin{table}[H]
\centering
\footnotesize
\caption{{\it StarberrySense} technical specifications}
\begin{tabular}{ll}
\hline
Weight (without baffle) & 315 gm \\
Weight (with baffle) & 470 gm \\
Board  &Raspberry Pi Zero \\
Dimensions (without baffle) (L$\times$W$\times$H) &$98 \times 75 \times 69$ mm \\
Dimensions (with baffle) (L$\times$W$\times$H) &$98 \times 75 \times 160$ mm \\
Power &1.25 W\\
Image Sensor & Sony IMX219 \\
FOV &$9.31^{\circ} \times 7^{\circ}$\\
Wavelength range & $450-750$ nm\\
Limiting magnitude & 6.5 (V)\\
Mode of operation & Lost-In-Space\\
Pointing Accuracy (3$\sigma$) & $27.18\as$\\
Roll Accuracy (3$\sigma$) & $38.76\am$ \\

\hline
\end{tabular}
\label{tab:specs}
\end{table}  

\subsection{Mechanical Structure and Assembly }

\begin{figure}[H]
    \centering
    \includegraphics[width= \linewidth]{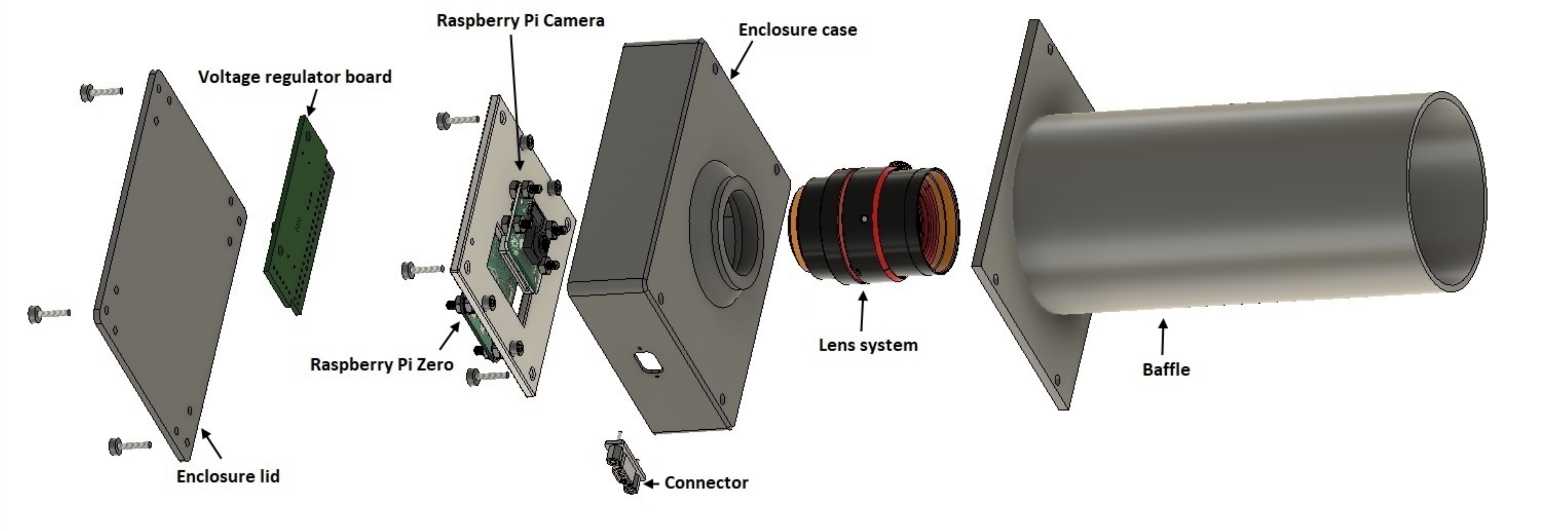}
    \caption{Exploded view of the mechanical structure.}
    \label{fig:model-2}
\end{figure}

The mechanical enclosure for the star sensor was custom-designed and precision-manufactured for holding the camera, C-mount lens, power supply and the RPi Zero. The enclosure was made with 3-mm thick Aluminium 6061. Figure~\ref{fig:model-2} shows the exploded view of the star sensor. The star sensor has a mass of about 315 gm, excluding the baffle. The entire housing is black anodized for better optical performance and to protect the surface from corrosion.  

\begin{figure}[H]
  \centering
  \includegraphics[width = 0.5\linewidth]{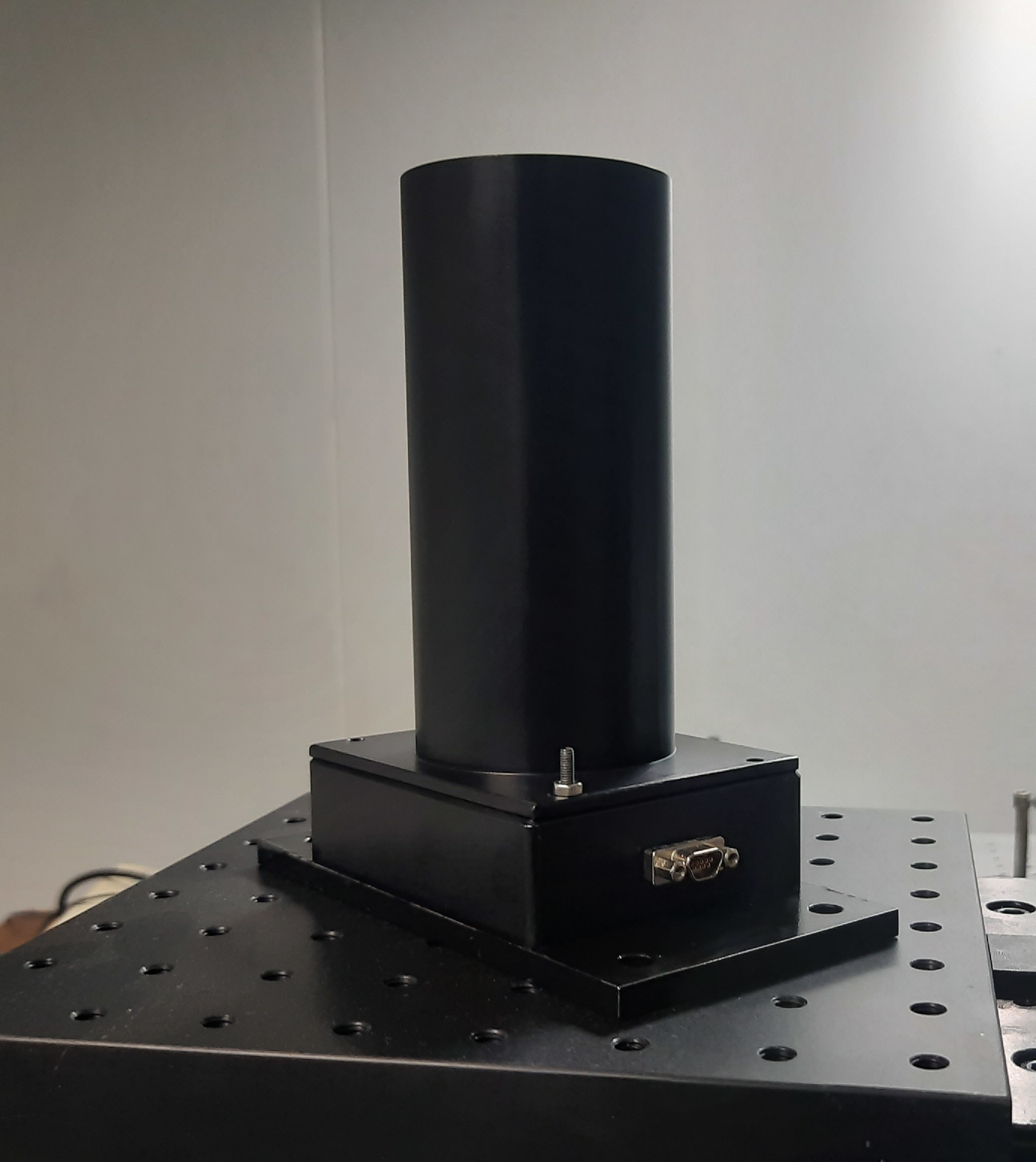}
  \caption{{\it StarberrySense} after assembly.}
  \label{fig:camera}
\end{figure}

To enhance the Sun and Moon avoidance angle, the star sensor requires a baffle. Otherwise, if the Sun appears near the FOV, it can saturate the detector, and the star sensor will not work. To prevent this, we designed a custom baffle following Asadnezhad \etal\cite{baffle} to achieve the Sun avoidance angle of $30\deg$, and manufactured it in-house from Aluminium 6061. The baffle was also black anodized to avoid scattering (Fig.~\ref{fig:camera}).

\begin{figure}[h!]
\begin{minipage}{.49\linewidth}
    \centering
    \includegraphics[width =\linewidth]{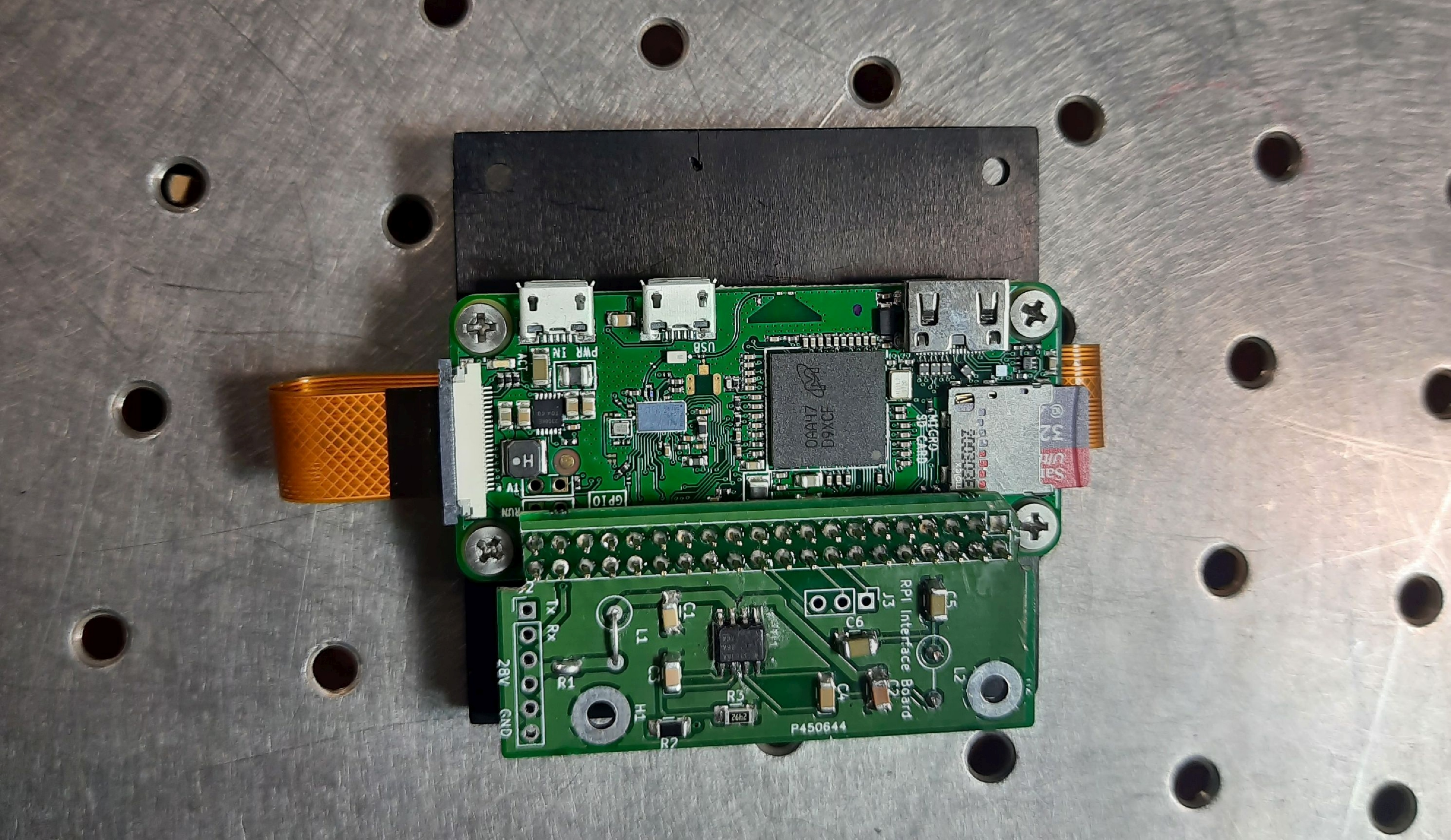}
    \captionof{figure}{Photo of the RPi-Zero and voltage regulator board mounted on the intermediate plate.}
    \label{rpi_1}
\end{minipage}\hfill
\begin{minipage}{.48\linewidth}
    \centering
    \includegraphics[width = .73\linewidth ]{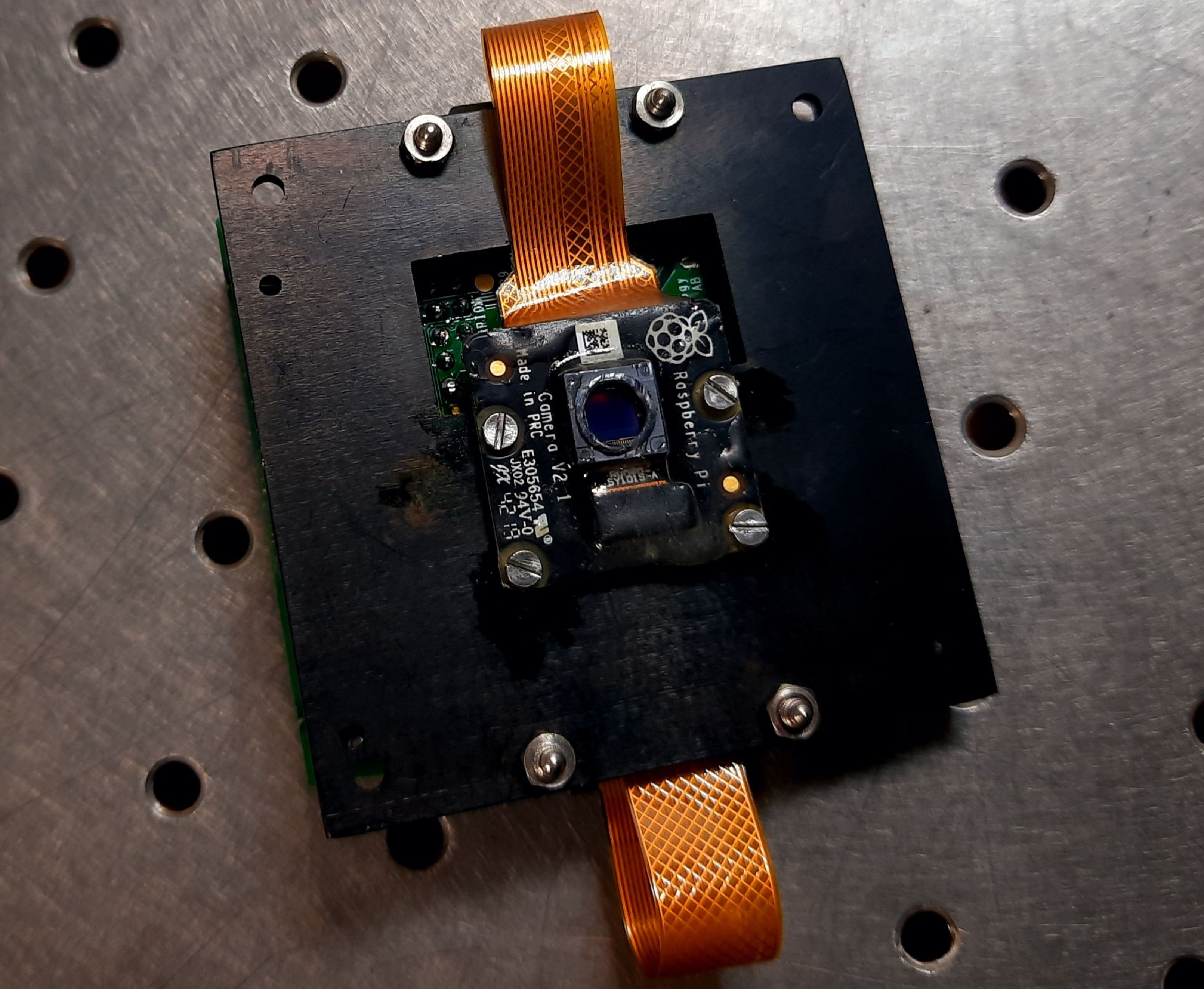}
    \captionof{figure}{Photo of the RPi v2 camera mounted on the intermediate plate.}
    \label{rpi_2}
\end{minipage}
\end{figure}

The assembly of the star sensor was initiated by bonding the lens mount for the C-mount lens system with the mechanical enclosure of the star sensor using epoxy Scotch-Weld 2216\texttrademark. Once the glue on the mount was cured, the C-mount lens assembly was threaded onto the mount. RPi camera and RPi Zero was mounted on an intermediate aluminium plate along with the voltage regulator board as shown in Figs.~\ref{rpi_1} and \ref{rpi_2}. The intermediate aluminium plate was mounted inside the star sensor enclosure by aligning it with respect to the optomechanical axis of the lens barrel and the RPi camera using a collimated light source. The alignment precision of the system was bound by the surface finish of the mechanical components of the star sensor, which was in the range of $50-100$ micron. Then, the lens system was moved on the C-mount thread to bring the focal plane onto the image sensor plane, and the lens system was locked in position by using the lock screw. Once the RPi camera was aligned, all the required electrical connections were made, and the enclosure was sealed using the enclosure lid. Once the main body of the entire star sensor unit was assembled, the baffle was fitted by using the mounting holes provided on the mounting plate of the enclosure case (Fig.~\ref{fig:camera}).   

\section{Software}

For the onboard operating system, we have used a minimalist version of Raspbian with in-house developed algorithms written in C++ running as a standalone application. For the image capture and the camera control, a C++ API developed by the AVA research group was used \cite{api}. The output of the sensor is attitude quaternions, which can be stored onboard and transmitted through the UART serial communication whenever needed.

\subsection{Lost in Space (LIS) Algorithm}

\begin{figure}[h]
     \centering
     \includegraphics[scale =0.8]{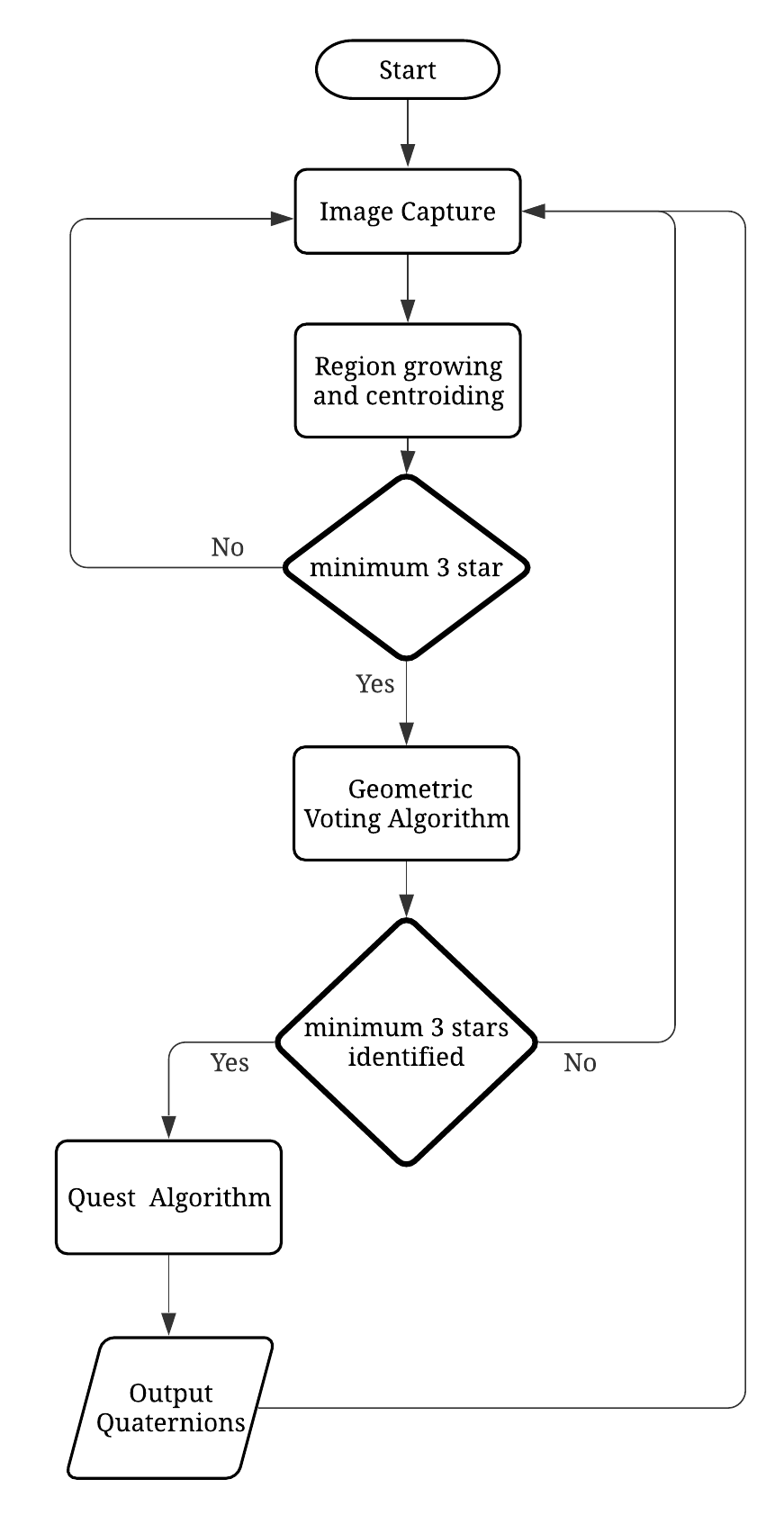}
     \caption{Flow chart of the algorithm onboard the star sensor}
     \label{flow_chart}
\end{figure}

The LIS algorithm is used to determine the orientation of the spacecraft in space without the need for any initial knowledge of the satellite orientation. The algorithm identifies stars in the FOV and determines the attitude of the satellite. The algorithm uses a search table for star identification, generated and stored onboard prior to launch. During the star sensor in-flight operation, the distance value between star pairs in the image is compared with the search table for the voting. Once the voting is completed, and the actual stars are identified, the QUEST algorithm is used to determine the orientation in terms of quaternions of rotation. Fig.~\ref{flow_chart} shows the flow chart of the algorithm implemented onboard the star sensor.

\subsubsection{Star Identification Process}

Several methods have been developed for the identification of stars in the images \cite{algorithms}. Initially, stars were identified based on their magnitudes which required a flux-calibrated image sensor. The first-star sensor, based on the CCD, was developed by Salomon at JPL \cite{1976}. The major flaw with this technique was that over time, as the sensor degraded, the method was prone to errors. Then came the development of the pattern-matching based algorithms \cite{algorithms} for star identification. Among them, one of the well-known methods was proposed by Liebe \etal \cite{star}, which was based on finding the angular distance from a star to its two closest neighbours and the spherical angle between them to identify the star. This method paved the way for the development of algorithms that use star triangles for star identification \cite{tri1,tri2,tri3,tri4,mortari-py,tri5}. Another well-known method, proposed by Mortari \etal \cite{mortari-py}, is the pyramid algorithm which uses four stars to perform the star identification. This technique is more robust and reliable.

In our star sensor, we have used the geometric voting algorithm with a binary search algorithm for star identification \cite{gv,mico}. The geometric voting uses angular distances from a star to all-stars in the FOV for voting to get the identity of the star. This is followed by the secondary voting to confirm these identities, where false stars are rejected, making it robust and less sensitive to noises. 

\subsubsection{Search Table Generation}

The search table is generated manually and stored onboard before the flight. The search table was generated by calculating the angular distances between all possible star pairs in the Hipparchus catalogue and saving only those star pairs whose angular separation is less than the longest FOV $D_{fov}$ (Table~2). Once the distances of these star pairs are calculated, they are sorted based on the increasing order of the distance between them and stored onboard the star sensor along with the unit vector of each star in the Earth-Centered inertial (ECI) coordinate system. The unit vectors of stars are calculated using the equation
\begin{equation}
\begin{bmatrix}
  \hat{x} \\
  \hat{y} \\
  \hat{z} 
\end{bmatrix}
=
\begin{bmatrix}
\cos{\alpha}\cos{\delta}  \\
\sin{\alpha}\cos{\delta}  \\
\sin{\delta}   
\end{bmatrix}\,,
\end{equation}
where $\alpha$ is Right Ascension, and $\delta$ is the declination of a star. Thus, the search table stored onboard contains the list of all possible star pairs along with the angular distances between them. Algorithm~\ref{code2} shown below was used to generate the search table, where $D_{fov}$ is the longest FOV and $d_{ij}$ is the minimum required angular separation between the stars, assumed to be $30^{\as}$ to avoid blending.

\begin{algorithm}
\footnotesize
\caption{Search table generation}
\label{code2}
\begin{algorithmic}[1]
\For {i=1 to N}
      \For {j=i+1 to N}
      \State $d_{ij}$ = distance between $N_{i}$ and $N_{j}$
      \If {$d_{ij} < D_{fov}$ and $d_{ij} > D_{min}$ }
       \State Append the table row with elements i,j,$d_{ij}$.
      \EndIf
      \EndFor
\EndFor
\State Sort the table based on $d_{ij}$.
\end{algorithmic}
\end{algorithm}

\subsubsection{Region Growing and Centroiding}

The first part of the algorithm onboard the star sensor is finding the centroids of the stars. For this, an image of the sky is acquired and run through a region-growing algorithm to grow the brightest regions in the image above a  threshold value. The threshold value is 
$p_{th}= \overline{p} + 5\sigma_{p}$, where $\overline{p} $ is the mean pixel value and $\sigma_{p}$ is the standard deviation in the pixel values. The pixels above a threshold that lies in the neighbourhood of 2 to 3 pixels are considered as part of the same region. Here single-pixel regions are ignored since they are mostly either due to cosmic ray events or random noise. After the regions are grown, the weighted centroids for all the regions in the image are calculated using the following formula,
\begin{equation}
     x = \frac{\sum_{i=0}^N x_i I_i}{\sum_{i=0}^N I_i}\,, \,\,  y = \frac{\sum_{i=0}^N y_i I_i}{\sum_{i=0}^N I_i}\,,
\end{equation}
where $(x,y)$ is the centroid, $(x_i,y_i)$ is the $i$th pixel position and $I$ is the intensity of the $i^{th}$ pixel. The weighted centroiding helps us achieve sub-pixel level accuracy. Once centroids are calculated, the unit vectors for each star in the camera reference coordinate system is calculated using the formula,
\begin{equation}
\begin{bmatrix}
  \hat{x} \\
  \hat{y} \\
  \hat{z} 
\end{bmatrix}
=
\begin{bmatrix}
\dfrac{pp_x \times (x-x_0)}{K \times f_{mm}}  \\
\dfrac{pp_y \times (y-y_0)}{K \times f_{mm}}  \\
\dfrac{1}{K}   
\end{bmatrix}\,,
\end{equation}
where {\footnotesize $$K= \sqrt{\left( (\dfrac{pp_x \times (x-x_0)}{ f_{mm}})^2 + (\dfrac{pp_y \times (y-y_0)}{ f_{mm}})^2 + 1\right)}\,.$$}.\\
Here, $f_{mm}$ is the focal length of the lens system, $(x_0,y_0)$ is the location of the central pixel of the detector, $pp_x$ is pixel size along $x$-axis and $pp_y$ is the pixel size along $z$-axis. Once the unit vectors for all the stars in the image have been calculated, geometric voting will begin.
 \begin{algorithm}[H]
\footnotesize
\caption{Star sensor algorithm}
\label{code1}
\begin{algorithmic}[1]
\State Calculate threshold for region growing  and perform region growing
\State Calculate the centroids of the stars along with uncertainty $\varepsilon$ of the centroid
\State Correct centroids for distortion
\State Assign \textbf{n} as the total number of centroids
\State Calculate $\vec{\textbf{p}}$, the unit vector for the centroids in sensor body coordinate
\For {i=1 to $n$}
  \For {j=i+1 to $n$-1}
    \State Compute distance $d_{ij} = \cos^{-1}(\vec{p_{i}}\cdot \vec{p_{j}})$
    \If{$d_{ij} < D_{fov}$}
      \State upper\_Index = binary\_search($d_{ij} + \varepsilon_{ij}$)
      \State
      lower\_Index = binary\_search($d_{ij} - \varepsilon_{ij}$) 
      \For{k = lower\_index to upper\_index}
       \For{all entries T(k)}
       \State Append voting list $V_i$ and $V_j$ of possible stars $S_i$ and $S_j$
      \EndFor
      \EndFor
    \EndIf  
\EndFor    
\EndFor
\For{i = 1 to n}
   \State Find the catalog star with maximum votes in voting list $V_{i}$
   \State Assign $St_{i}$ to the catalogue star which got the maximal votes
\EndFor
\For{i = 1 to n}
   \For{j = i+1 to n}
     \State Compute distance $d_{ij} = \cos^{-1}(\vec{p_{i}}\cdot \vec{p_{j}})$
     \If {distance between stars $St_i$ and $St_j \in d_{ij}$} 
          \State Add a vote for the match of ($St_i$,$S_{i}$) and ($St_j$,$S_{j}$) 
     \EndIf
    \EndFor
\EndFor
\State Identify true stars based on primary and secondary votes.
\State Estimate attitude using QUEST algorithm.
\end{algorithmic}
\end{algorithm}
Many factors cause the measured centroids to shift from their actual positions. Centroiding accuracy is proportional to the square root of the signal from the stars, and if the point-spread function (PSF) is less than 0.5 pixels, the centroiding accuracy will be limited by sampling theorem \cite{accu-per}. Another factor is the radial distortion caused by the lens system. This results in the centroids being displaced radially. This distortion must be accounted for during the calibration process with a suitable distortion model and corrected during the processing \cite{gv,accu-per} (Sec.~4.2).

\subsubsection{Geometric Voting }

In geometric voting, the distance between each star pair is calculated for the matching process. The distance can be calculated by finding the inverse cosine of the unit vector dot product of the star pairs. Once the distance is found, the star's ID should be in the search table in the distance range from $d_{ij} - \varepsilon $ to $d_{ij} + \varepsilon$, where $\varepsilon$ is the uncertainty in calculated distance \cite{gv}, assumed to $20^{\as}$ in our case. Using a binary search algorithm, the location on the search table is found where this distance range lies. Once the location of the distance range is known, the stars IDs, lying between the distance $d_{ij} - \varepsilon$ and $d_{ij} + \varepsilon$, will be used for the voting. After the voting, the total votes are counted to identify the star in the image that corresponds to the star in our catalogue. The second round of voting is used to remove falsely identified stars. In the end, the total number of votes from primary and secondary rounds is used to identify the stars and to remove false detections. Once the stars in the image are identified, the star sensor proceeds to the attitude determination.

\subsection{Attitude Determination}
\begin{figure}[H]
    \centering
    \includegraphics[scale =0.5]{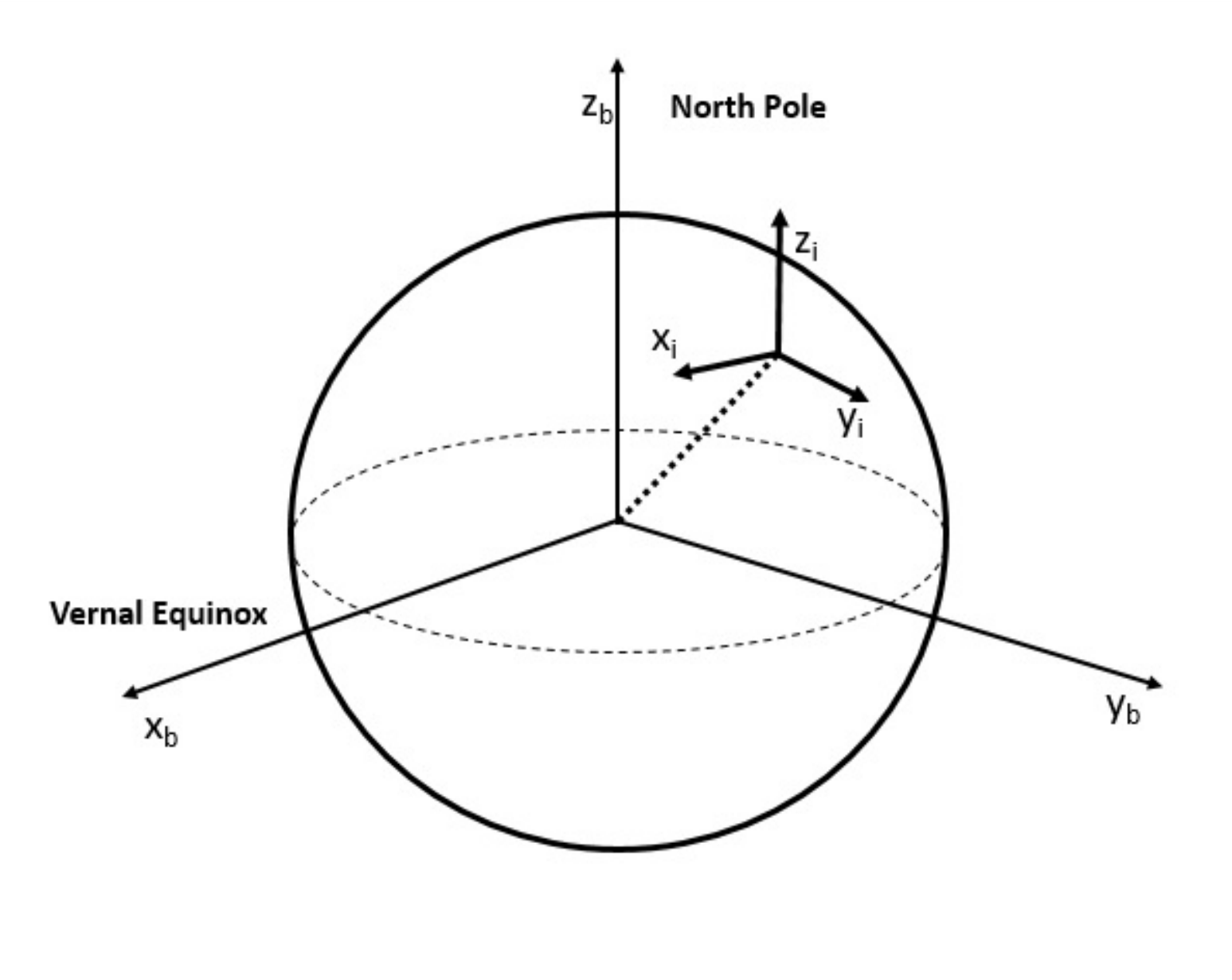}
    \caption{Figure showing the ECI-coordinate system and the sensor body coordinate system}
    \label{corrdinates}
\end{figure}

Algorithm \ref{code1} is used to identify stars and determine the attitude. Once stars in the image are identified, the final step is to calculate the quaternion of rotation between the ECI coordinate system and the sensor body coordinate system. The geometric voting algorithm provides the unit vector of the identified star in the ECI coordinate system, and the unit vector for that star in the sensor coordinate system is calculated using Eq.~(2). 

Two commonly used algorithms to determine the attitude from the unit vectors are TRIAD\cite{triad} and QUEST\cite{quest}. Tri-Axial Attitude Determination (TRIAD) determines the direction cosine matrix that describes the relationship between two coordinate systems. TRIAD algorithm only uses two of the unit vector pairs and discards the rest, thus not utilizing the complete information from all the star unit vector pairs. Since each identified star has a pair of unit vectors, the total amount of pairs can be 20 or more (see Fig.~\ref{rs2} for an example of 18 registered stars). Since we want to use the complete information, the algorithms suited for handling such cases are Davenport's q-method\cite{q-method}, QUEST and SVD\cite{svd, thesis}. Since QUEST bypasses the eigenvalue problem and is computationally less expensive, we have implemented it in our attitude determination algorithm.

Using the QUEST algorithm with multiple unit vector pairs as inputs, the star sensor calculates the quaternion of rotation between the two coordinate systems. The rotation can be described in many ways, such as rotation matrix, Euler angles, quaternions, etc. In our case, we have chosen quaternions as they are computationally less intensive. Now, if $v_b$ is the unit vector for the star in the sensor body coordinate system and $v_i$ is the unit vector for the ECI coordinate system
(Fig.~\ref{corrdinates}), we have
\begin{equation}
v_b=R v_i\,,
\end{equation}

where $R$ is the rotation matrix that describes the rotation between the ECI and the sensor body coordinate systems. Once the stars in the image are identified, we know the unit vector of those stars in both coordinate systems. Now, the star sensor needs to find the rotation matrix $R$ from the unit vectors. The value of $R$ must be such that it minimises the loss function 
%
\begin{equation}
     J= \frac{1}{2} \sum_{k=1}^{N}  w_k |v_{kb} - R v_{ki}  |^2  \,,
\end{equation}
where $J$ is the loss function, $w_k$ is the weight of each star unit vector pair, and $N$ is the number of correctly identified stars. Since the measurements are not ideal, we will always have a value of $J$ greater than zero. The star sensor needs to find a solution that minimises the loss function $J$, and the method should not be computationally intensive. There are several existing methods to solve this classic Wahba’s problem \cite{wahba}. We restate the loss function in terms of quaternions in such a way that it becomes an eigenvalue problem, where the largest eigenvalue is to be found. The QUEST algorithm was developed to bypass the expensive eigenvalue problem by approximating the process. It reduces to solving for parameter $p$, called the Rodriguez parameter, in the equation
\begin{equation}
    \left[\left(\lambda_{opt}- \sigma\right)I-S\right]p=Z \,,
\end{equation}
where 
\begin{gather*}
  Z = \left[B_{23}-B_{32}B_{31}-B_{13}B_{12}-B_{21}^T\right]\,,\\
B=\sum_{k=1}^{N}w_k\left(v_{kb}v_{ki}^T\right)\,,\\
\lambda_{opt}=\sum w_k\,,\\
\sigma =Tr(B)\,,
\end{gather*}
and
\begin{equation}
  S=B+B^T\,.
\end{equation}
After finding the value of $p$, the attitude quaternion is given by
\begin{equation}
    q_{quest}= \frac{1}{\sqrt{1+p^T p}} \begin{bmatrix}
p\\
1
\end{bmatrix}\,,
\end{equation}
where
\begin{equation}
   q_{quest}=\begin{bmatrix}
q_1\\
q_2\\
q_3\\
q_4
\end{bmatrix}\,. 
\end{equation}

\section{Calibration and Testing}

\subsection{Performance and precision limit}

We calculated the theoretical limit of accuracy for noise equivalent angle (NEA) following calculations in Liebe (2002) \cite{1008988}. NEA measures the ability of a star sensor to reproduce the same attitude information for the same part of the sky. Even though NEA depends on different instrumental parameters such as dark noise, read noise, etc., it is possible to quickly estimate it. For that, we first find the average number of stars in the FOV $N_{stars}$,
\begin{equation}
    N_{stars} = N_{catalog}\frac{1-\cos{\frac{D}{2}}}{2}=11\,,
\end{equation}
where $N_{catalog}$=8874 is the total number of stars from the Hipparchus catalog used to generate the search table (stars with magnitudes  less than or equal to 6.5). The average {\it StarberrySense} FOV is $D=8.15\deg$ and the average number of pixels is $N_{pixels} $=1436. 

The cross boresight NEA ($E_{CB}$), which gives the bound of error, or uncertainty, in the pointing direction; and the roll axis NEA ($E_{Roll}$), that gives the bound of error, or uncertainty, in the roll axis, are calculated as
\begin{equation}
     E_{CB} = \frac{D E_{centroid}}{N_{pixels} \sqrt{N_{stars}}} = 3.05^{\as}\,,
\end{equation}
and
\begin{equation}
     E_{Roll} = \tan^{-1}\left({\frac{E_{centroid}}{0.3825N_{pixels}}}\right) \frac{1}{\sqrt{N_{stars}}}=  56.06^{\as} \,,
\end{equation}
where $E_{centroid}$ is the centroiding accuracy of the star sensor, assumed to be 0.5 pixels as the worst-case scenario. 

\subsection{Real Sky Test and Distortion Correction} 

After the star sensor was assembled, a night-sky test was conducted at Vainu Bappu Observatory (VBO) facility of IIA, Kavalur, Tamil Nadu. The star sensor was mounted on a stable platform and pointed towards the night sky with no pointing knowledge. Two subsequent night-sky tests were conducted, the first to correct for the distortion of the lens system and the second to verify the pointing performance. 

The main distortion that affects the angular measurement of the distance between stars is the radial distortion which causes a significant error during the voting process leading to false star identification. The obtained image of the sky in jpeg format was fed to Astrometry.net \cite{astrometry} for astrometric calibration and generation of distortion-correction polynomials coefficients. Thereafter, these polynomials coefficients were stored onboard the star sensor for performing distortion correction using the SIP convention method \cite{sip}. 
If $(u,v)$ is the centroid of the star in the pixel coordinate system and $(x,y)$ is the centroid of the star after the correction, we have the distortion correction equation
\begin{align}
   x & = u +  \sum_{a,b} A\_a\_b \ u^{a} v^{b}, \nonumber\\
   y &= v +  \sum_{a,b} B\_a\_b \ u^{a} v^{b}, 
\end{align}
where $A\_a\_b$ and $B\_a\_b$ are polynomial coefficients of the second order, and 
$a$ and $b$ are integers, $a + b \le 2$. This correction was implemented in the Algorithm \ref{code1} after the centroid calculation, and the corrected centroids were used for unit vector calculation. We found that second-order polynomial correction was sufficient as the distortion was less than a $26^{\as}$ even at the edges.
\begin{figure}[ht]
    \centering
    \includegraphics[width = 0.7\linewidth]{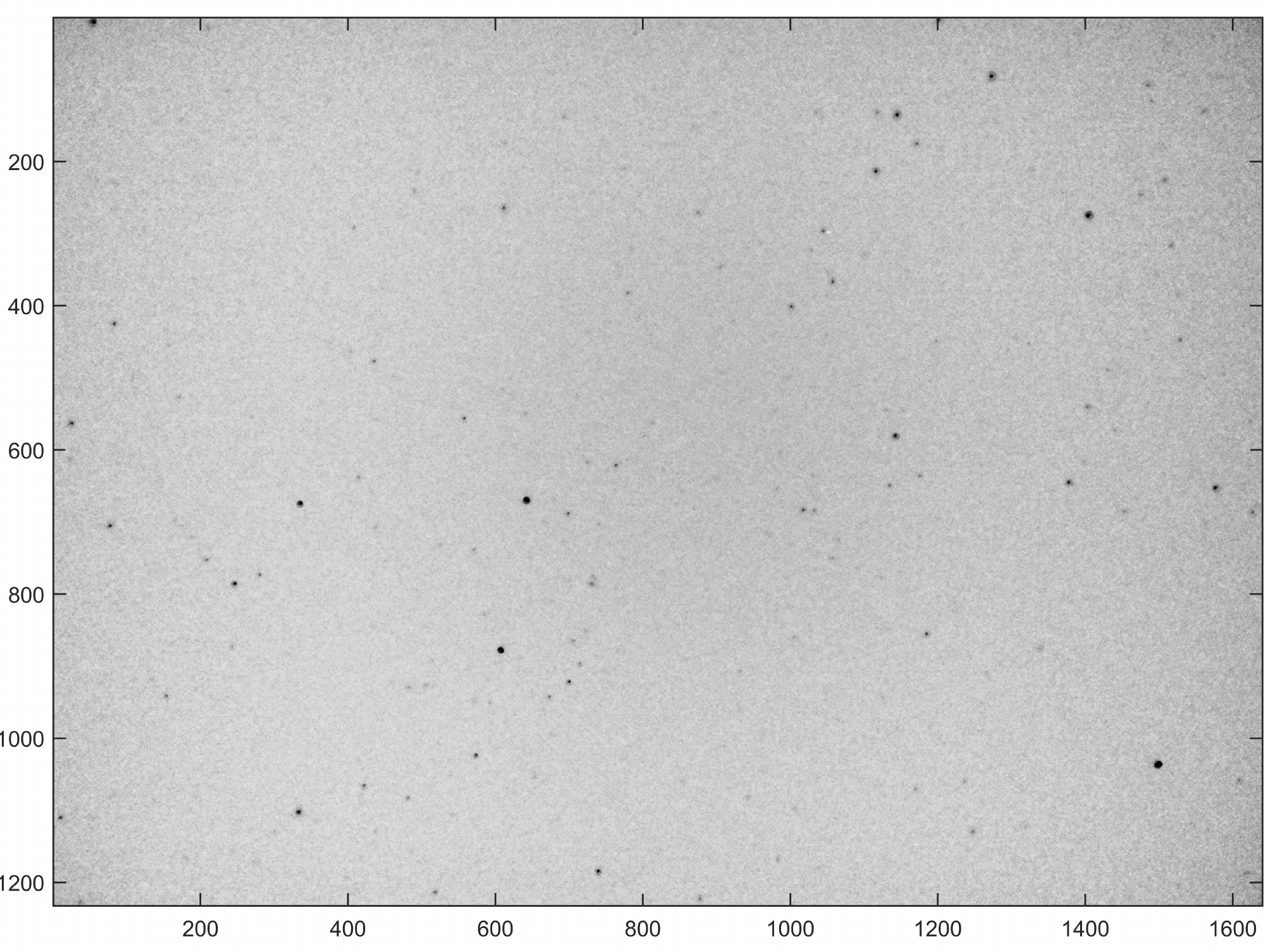}
    \caption{500-ms exposure inverted grey-scale image of the night sky captured with the star sensor. $x$ and $y$ axes are pixel numbers.}
    \label{rs1}
\end{figure}

\begin{figure}[ht]
    \centering
    \includegraphics[width = 0.7\linewidth]{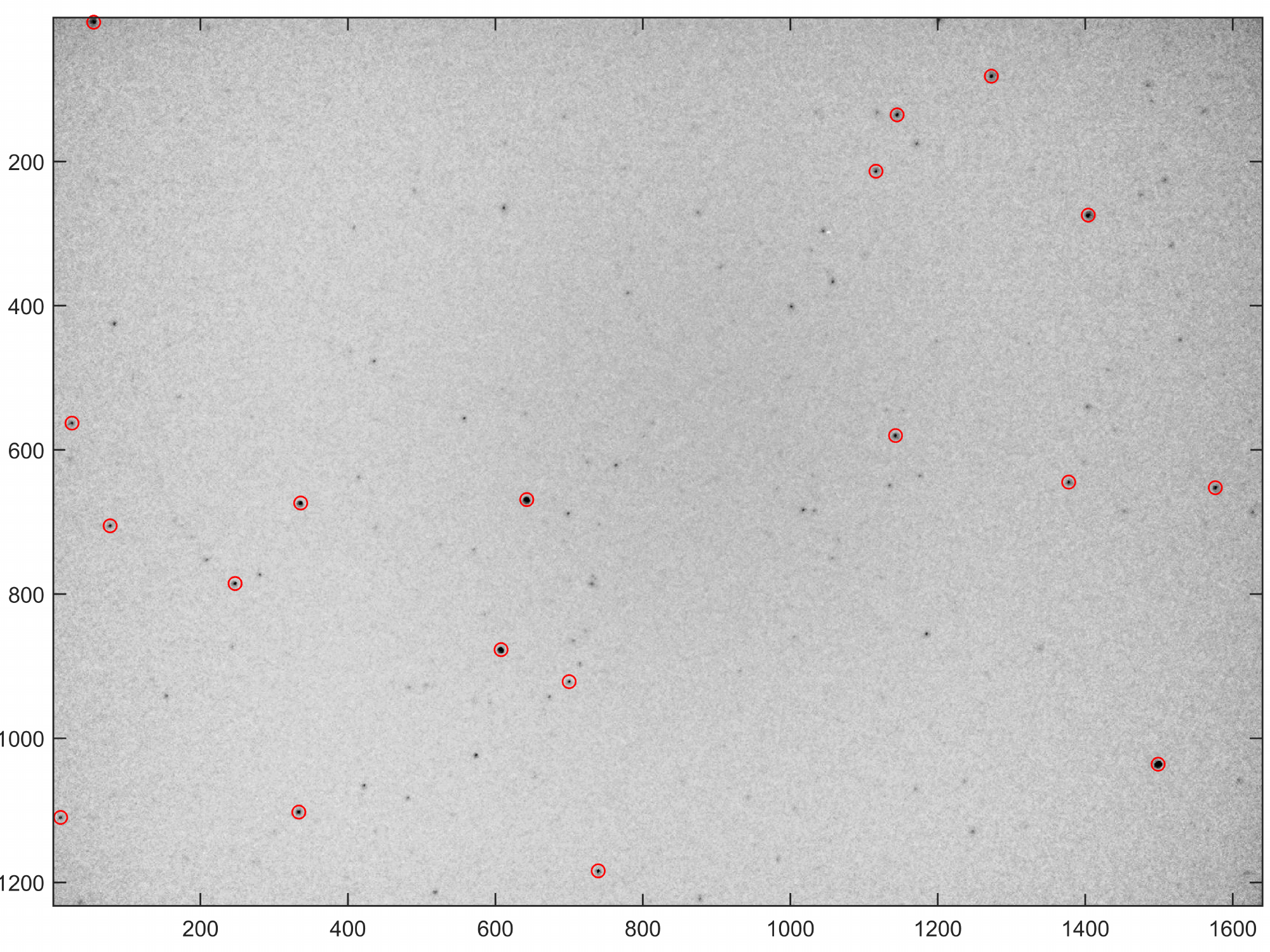}
    \caption{Same image as Fig.~9 with marked centroids of the stars above the threshold (Sec.~3.1.3) after running the region-growing algorithm . $x$ and $y$ axes are pixel numbers.}
    \label{rs2}
\end{figure}
In the second test, the star sensor was pointed at different parts of the sky, and the captured images were stored onboard along with calculated quaternions. Fig.~\ref{rs1} shows one of the captured images, pointed at a random part of the sky, and the calculated centroids of the stars after running the region-growing algorithm are shown in Fig.~\ref{rs2}. Quaternions, obtained from the star sensor, were used to find the pointing \cite{markley2014fundamentals}, which in case of Fig.~\ref{rs1} was $\alpha =91.0176\deg$ and $\delta= 14.1498\deg$, while the actual pointing from Astrometry.net was $\alpha =91.019\deg$ and $\delta=14.149\deg$. The ground-based observations error was found to be $6^{\as}$ when pointing from the star sensor was compared with the pointing obtained from Astrometry.net.

\begin{figure}[ht]
\begin{minipage}{.45\linewidth}
   \centering
  \includegraphics[width = \linewidth]{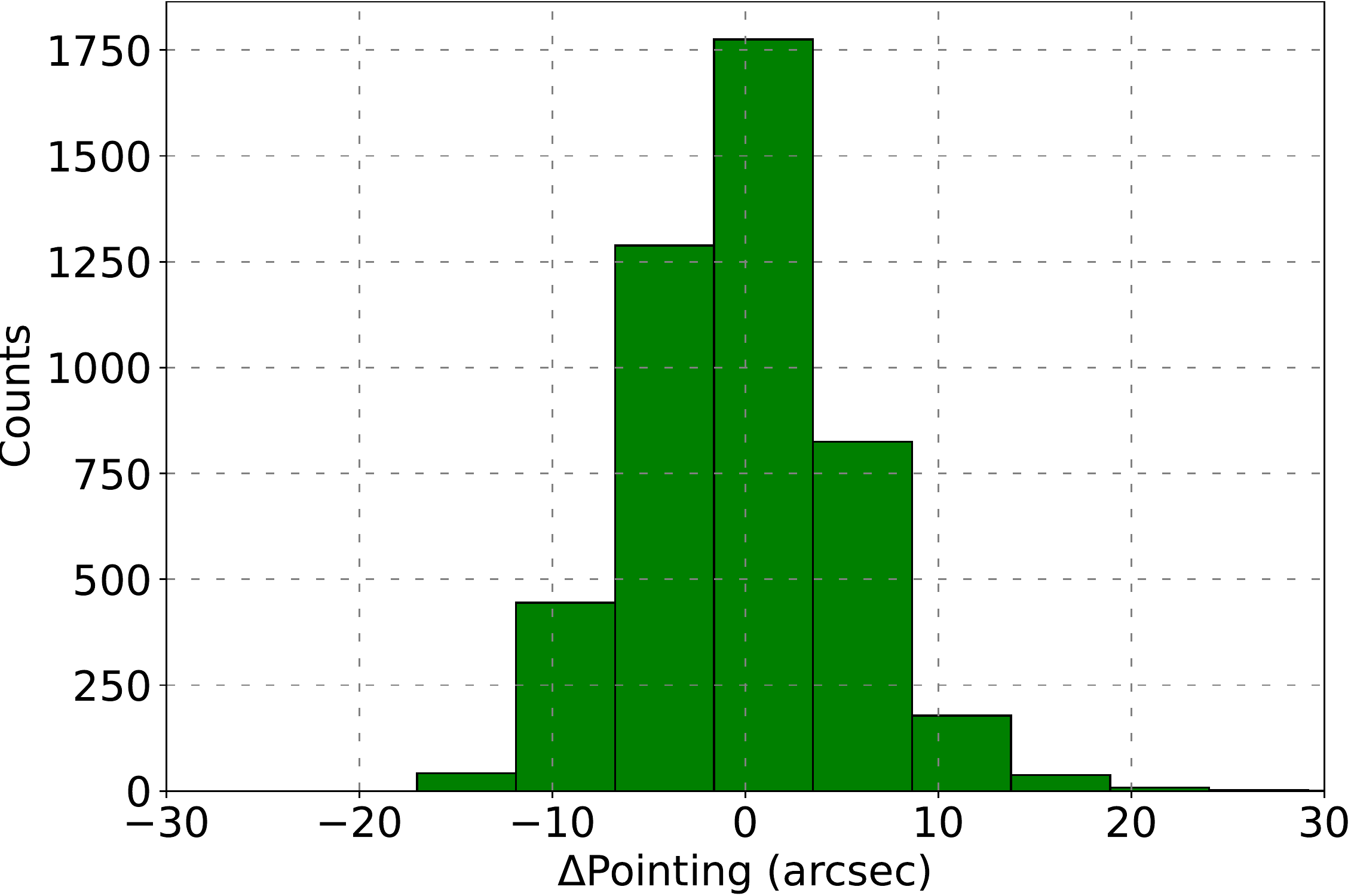}
  \caption{ Plot shows the histogram of error for the {\it StarberrySense} in the pointing axis. }
  \label{fig:pointing}
\end{minipage}\hfill
\begin{minipage}{.45\linewidth}
\centering
  \includegraphics[width = \linewidth]{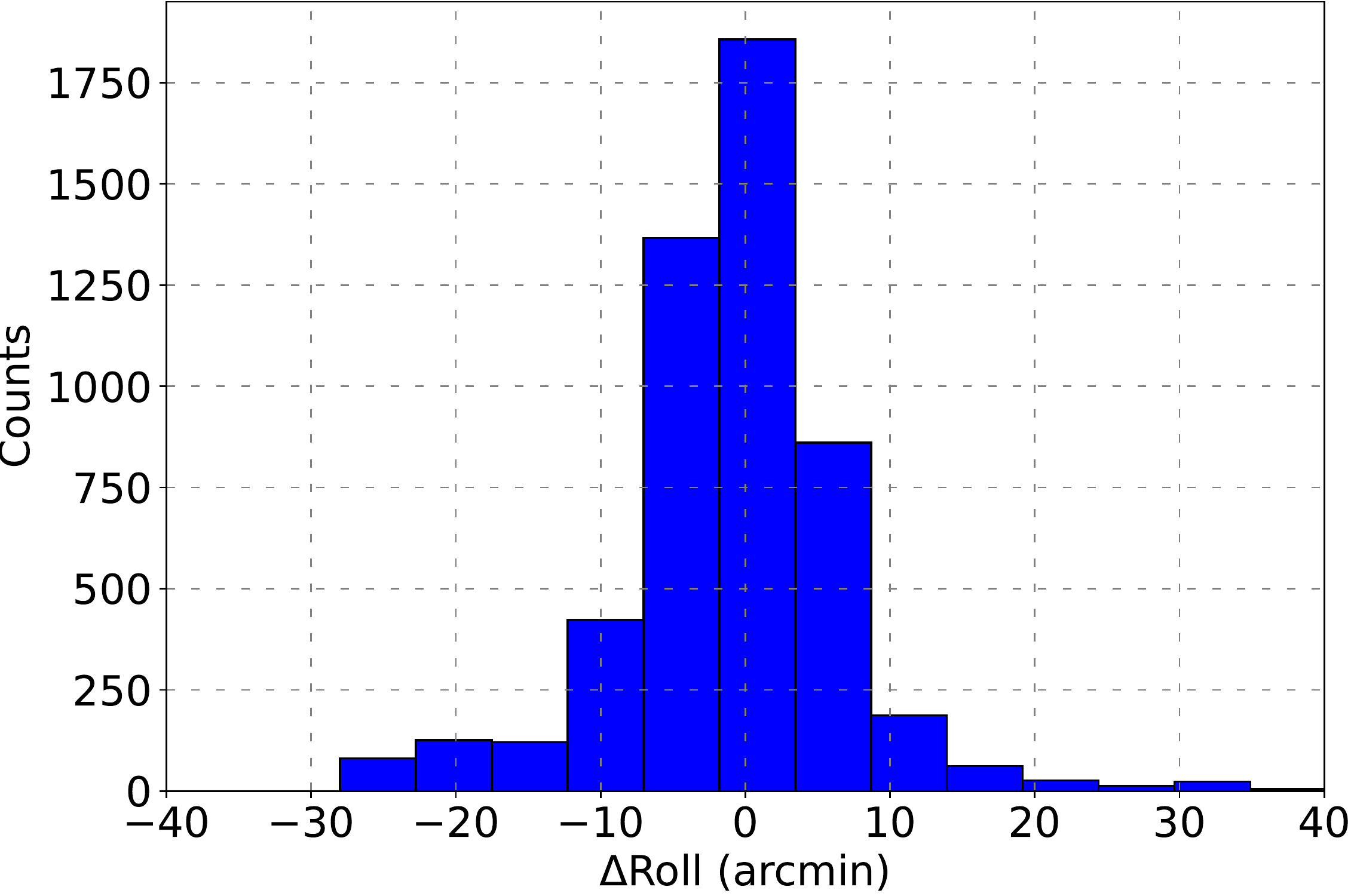}
  \caption{ Plot shows the histogram of error for the {\it StarberrySense} in the roll axis.}
  \label{fig:roll}
\end{minipage}
\end{figure}
To determine the accuracy of the {\it StarberrySense} in the pointing and roll direction, the star sensor was mounted on a stable platform and pointed towards the sky. The star sensor was set to calculate the quaternions and store them on board along with the sky images for a long duration. The actual pointing obtained from the Astrometry.Net using the stored images and the pointings obtained from the star sensor was used to determine the accuracy in the pointing and in the roll axis. Fig.~\ref{fig:pointing} and Fig.~\ref{fig:roll} show the histogram of error in the pointing and roll axis, respectively. The measured accuracy in the pointing axis was $3\sigma$  = 27.18 arcsec, and in the roll axis was $3\sigma$= 38.76 arcmin. The success rate of the star sensor was found to be 94.69\% from the real-sky test. Also, the sky condition during the test was close to 5 on the Bortle scale, which is a measure of the night sky brightness \cite{bortle}. We also determined the slew rate limit by simulating the star trailing on the images captured by the star sensor. We found that after the simulated motion rate at 0.2$\deg$/s, the algorithm failed to determine the quaternions.
\subsection{Power Consumption and Processing Time}

Table \ref{tab:power} shows the average power consumption during different processes: when the star sensor is idle, capturing the image, processing the image, or during UART transmission. The star sensor draws an average current of 140 mA when idle, and the current can peak to 220 mA during the image capture.
\begin{table}[H]
\centering
\footnotesize
\caption{Star sensor power consumption.}
\begin{tabular}{ll}
\hline
Process & Power \\
\hline
Idle & $ 0.7$ Watts \\
Image capture and readout & $1.1$ Watts \\
Image Processing & $0.9$ Watts \\
UART transmission & $0.75$ Watts\\
\hline
\end{tabular}
\label{tab:power}
\end{table}
Table~\ref{tab:speed} shows the processing time taken by each of the subroutines in the star sensor algorithm. The image capture and readout take the longest amount of time, and the star sensor can provide 14 quaternions per minute. 

\begin{table}[H]
\centering
\footnotesize
\caption{Star sensor processing time for the different subroutines.}
\begin{tabular}{ll}
\hline
Process  & Time \\
\hline
Image capture and readout & $3.6$ s \\
Threshold determination & $240$ ms \\
Centroiding & $220$ ms \\
Geometric Voting & $50$ ms \\
QUEST  & $50$ ms\\
\hline
\end{tabular}
\label{tab:speed}
\end{table}

\section{Flight Qualification}

To qualify for the space flight, payloads have to undergo the standard environmental tests: vibration and thermal-vacuum tests. This is to ensure that the payload will be able to withstand all launch loads and operate in the space environment. The most stringent vibration requirements for launch vehicle platforms are the following: the natural frequency of the payload must be above 100 Hz, and the instrument should be able to withstand acceleration loads up to 10g\cite{Mathew_2017, Kumar_UVIT_2012}. The thermal-vacuum and vibration tests for the sensor were performed in the M.~G.~K.~Menon Lab, CREST Campus, IIA, Bangalore, as per the Polar Satellite Launch Vehicle (PSLV) Stage 4 flight requirements.

\subsection{Thermal-vacuum test}

\begin{figure}[h]
\centering
\begin{minipage}{.54\textwidth}
    \centering
    \includegraphics[height = 2.3in]{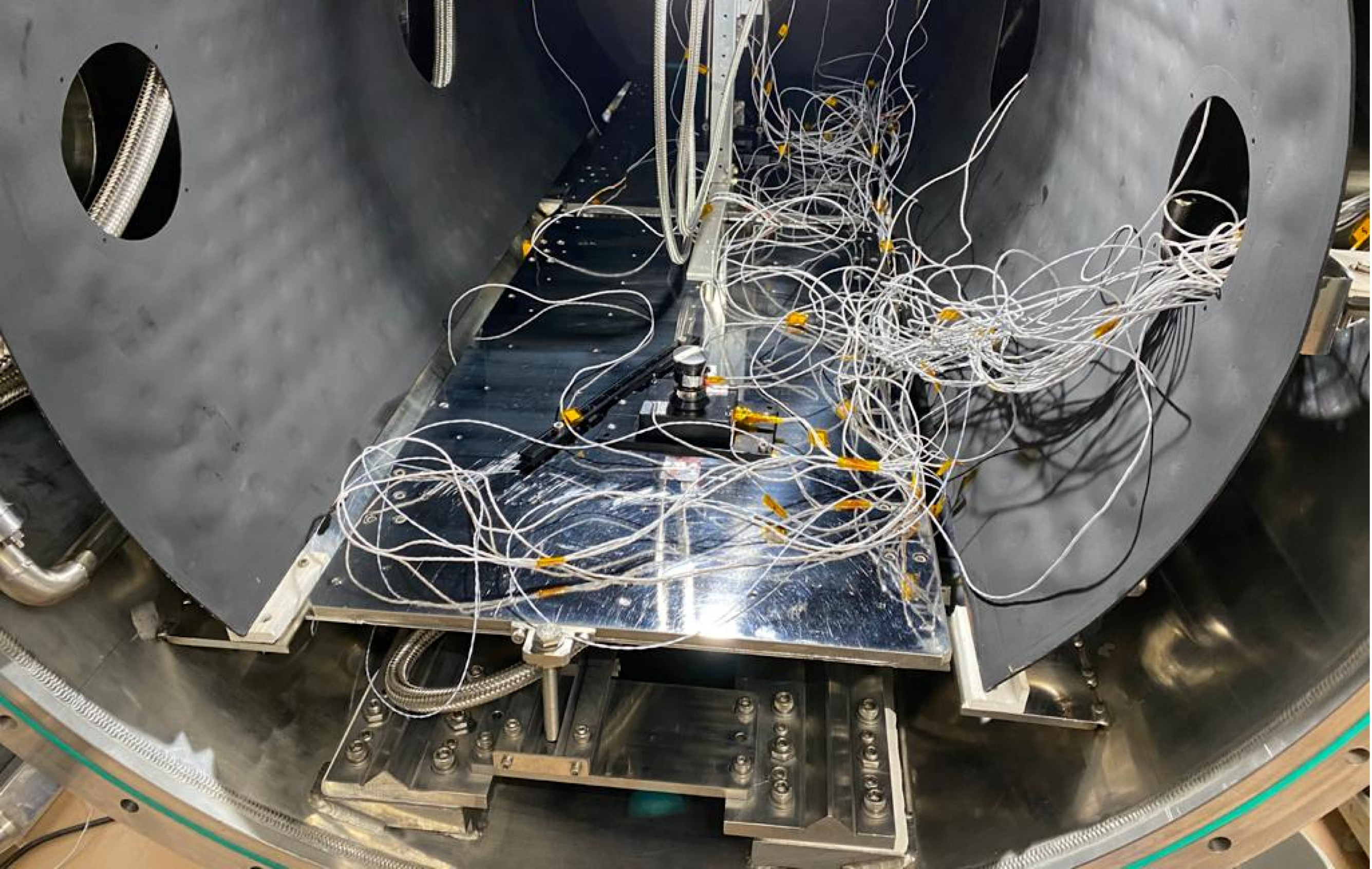}
    \captionof{figure}{Star sensor mounted inside the thermal-vacuum chamber in MGKM lab, CREST, IIA.}
    \label{thermo_im}
\end{minipage}\hfill
\begin{minipage}{.43\textwidth}
    \centering
    \includegraphics[height =2.3in]{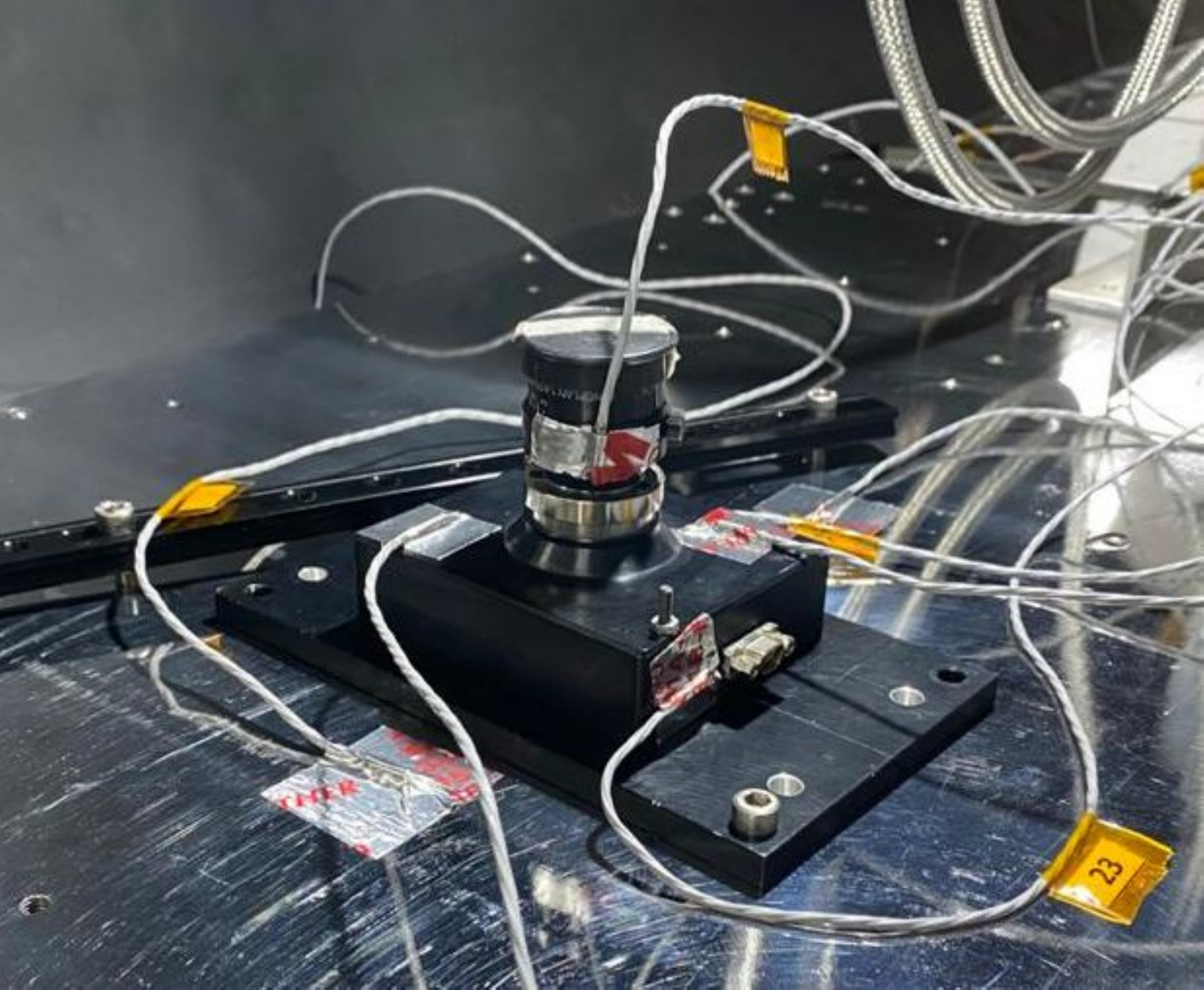}
    \captionof{figure}{Star sensor with temperature sensors taped on with aluminium tape.}
    \label{thermo_sensor}
\end{minipage}
\end{figure}

\begin{figure}[h]
    \centering
    \includegraphics[trim={2cm 0 2cm 0},width=1\linewidth]{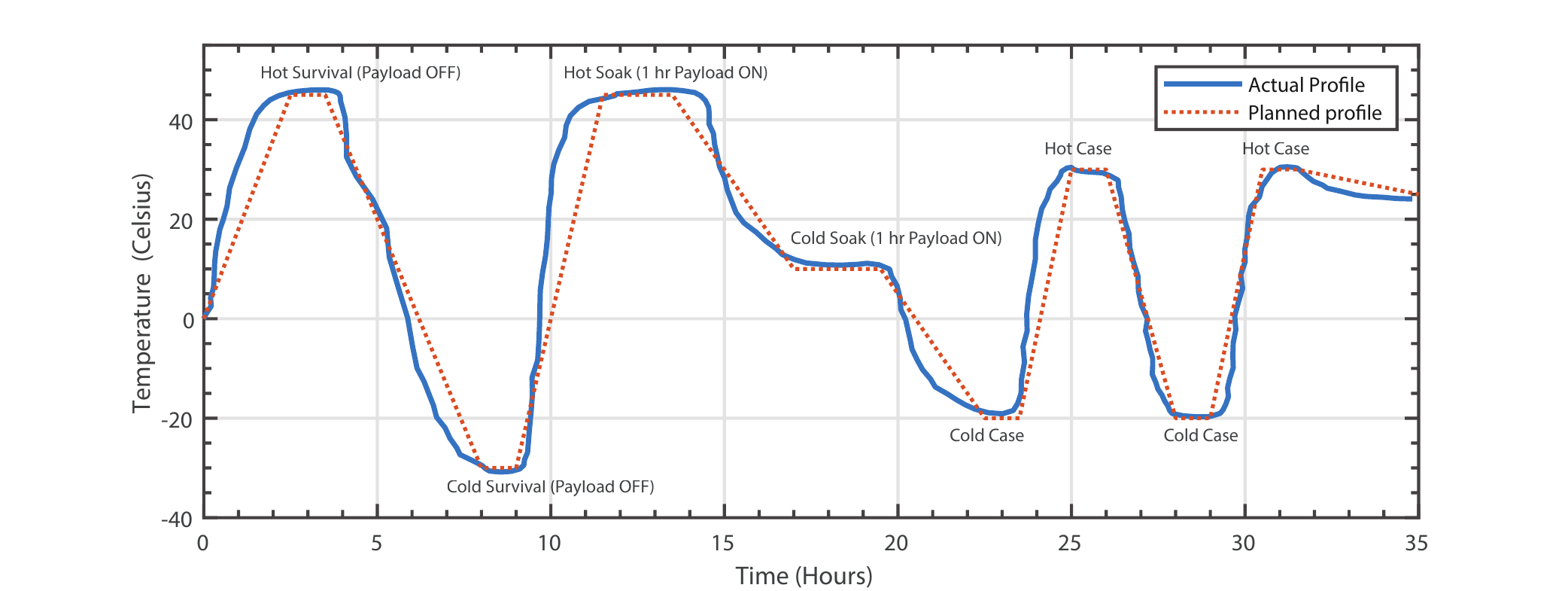}
    \caption{Plot shows the average temperature values recorded from the four sensors mounted on star sensor body during the thermal-vacuum test (Blue continuous) overplotted on the programmed temperature profile (Red dotted).}
    \label{thermoac}
\end{figure}

A thermal-vacuum chamber was used to simulate the thermal environment that the sensor will be subjected to in space. Four temperature probes were mounted on the {\it StarberrySense} body as shown in Figs.\ref{thermo_im} and \ref{thermo_sensor} using aluminium tape, then the pressure inside the chamber was pumped down to $10^{-6}\, mbars$ with the help of a roughing pump and a turbo-molecular pump. The temperature in the chamber was varied through seven cold and hot cycles to simulate the orbital environment in the LEO. Fig.~\ref{thermoac} shows the temperature measurement from the four sensors mounted on the star sensor overplotted on the programmed chamber profile\cite{thermovac}.

  After each hot and cold cycle, the functionality of the payload was checked by turning on the system and capturing dark frames, which were stored onboard the system memory. The camera module and RPi functioned without any errors, and the dark frames were consistent without any abnormalities. Thus, the payload remained functional and was able to handle the temperature variations.

\subsection{Vibration test}

\begin{figure}[H]
    \centering
    \includegraphics[width =0.9\textwidth]{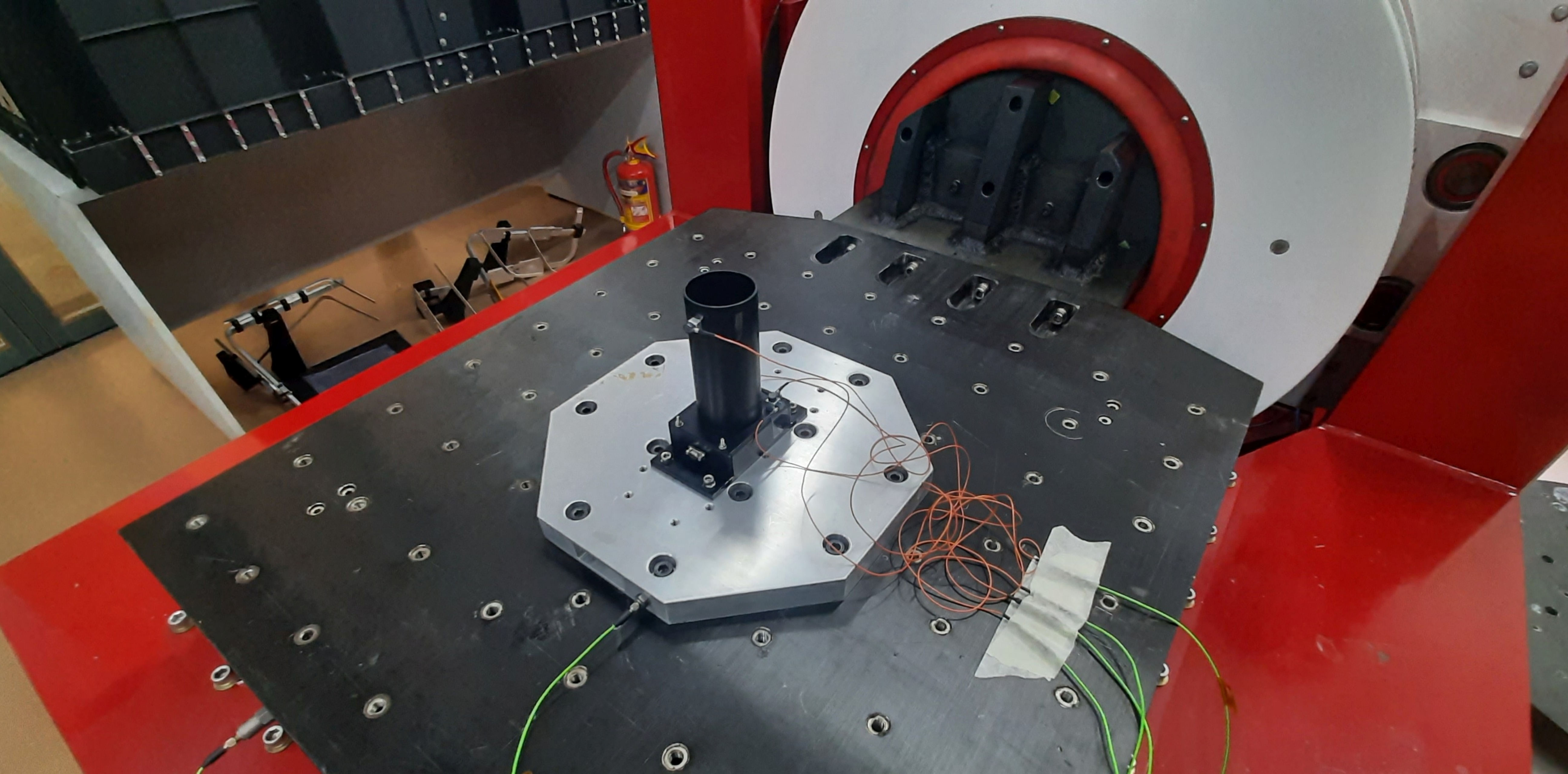}
    \caption{ {\it StarberrySense}  mounted on the vibration table with accelerometers attached to the body in the MGKM lab, CREST, IIA.}
    \label{vib_setup}
\end{figure}

 The vibration test was conducted as per the requirement of the launch provider (ISRO-PSLV). The launch load profiles are shown in Tables~\ref{sine} and \ref{random}. The payload was subjected to sine and random vibrations according to the launch requirements. Accelerometers were mounted on the {\it StarberrySense} body as shown in Fig.~\ref{vib_setup}. The {\it StarberrySense} successfully passed the vibration test without any structural damage, with all the connecting bolts remaining intact. The first natural frequency was found to be 385 Hz (well above the frequency level requirement of PSLV stage 4), and the next five successive frequencies were 585Hz, 911Hz, 1327Hz, 1376Hz and 1655Hz, respectively. After the vibration test, the star sensor was tested in the open sky. We verified that the star sensor operates normally as expected, and distortion parameters were unchanged.

\begin{table}[h!]
\centering
\begin{tabular}{|l|l|l|l|l|}
\hline
Frequency (Hz) & PSD ($g^2$/Hz) & Level                       & Duration               & Axis                            \\ \hline
20             & 0.001          & \multirow{5}{*}{9$g_{rms}$} & \multirow{5}{*}{1 min} & \multirow{5}{*}{All three axis} \\ \cline{1-2}
68             & 0.001          &                             &                        &                                 \\ \cline{1-2}
250            & 0.062          &                             &                        &                                 \\ \cline{1-2}
1000           & 0.062          &                             &                        &                                 \\ \cline{1-2}
2000           & 0.015          &                             &                        &                                 \\ \hline
\end{tabular}
\caption{Random vibration specifications for {\it StarberrySense} }
\label{sine}
\end{table}

\begin{table}[]
\centering
\begin{tabular}{|l|ll|ll|}
\hline
\multicolumn{1}{|c|}{\multirow{2}{*}{Frequency}} & \multicolumn{2}{l|}{Longitudinal axis}    & \multicolumn{2}{l|}{Lateral axis}         \\ \cline{2-5} 
\multicolumn{1}{|c|}{}                           & \multicolumn{1}{l|}{Level}   & Sweep rate & \multicolumn{1}{l|}{Level}   & Sweep rate \\ \hline
10 Hz - 16 Hz                                    & \multicolumn{1}{l|}{20mm DA} & 2 oct/min  & \multicolumn{1}{l|}{15mm DA} & 2 oct/min  \\ \hline
16 Hz - 100 Hz                                   & \multicolumn{1}{l|}{10g}     & 2 oct/min  & \multicolumn{1}{l|}{6g}      & 2 oct/min  \\ \hline
\end{tabular}
\caption{Sinusoidal specifications for {\it StarberrySense} }
\label{random}
\end{table}

\section{Conclusions and Future Work}

In this paper, we have presented the design and development of {\it StarberrySense}, a low-cost star sensor based on Raspberry Pi Zero. The star sensor was assembled, aligned, calibrated and tested in-house. A computationally-efficient algorithm has been implemented and tested against the real sky conditions, and the results were favourably compared with obtained values from Astrometry.net. The thermal-vacuum test for the star sensor was completed successfully, along with the vibration test as per PSLV launch requirements. The {\it StarberrySense} was selected by the recent Announcement of Opportunity call for payloads to fly on the PS4-Orbital Platform by ISRO. PS4-Orbital Platform is an innovative use of the spent PS4 stage (fourth stage of the PSLV) as it has standard interfaces/packages for power and telemetry and can be stabilized. Thus, it can carry the scientific payloads to perform scientific experiments for up to 6 months in LEO. We are preparing the payload for flight before the end of 2022 on board the ISRO PSLV stage-4 platform.\\
Further improvements in the next version of the {\it StarberrySense} include an up-gradation of the hardware processing capability with the newer Rpi Zero 2 as well as reduction in form factor and weight of the payload. In addition to that, an improvement in the optics is envisaged to enhance the light collection area and thereby reducing the exposure time.

\section{Acknowledgement}

We wish to thank Mr. S.~Sriram of the Indian Institute of Astrophysics (IIA) and Mr. Ajin Prakash from Arksa Research Labs, Bangalore, India, for their valuable suggestions and help. MS acknowledges the financial support by the Department of Science and Technology (DST), Government of India, under the Women Scientist Scheme (PH), project reference number SR/WOS-A/PM-17/2019. We also  thank all the staff at the M.~G.~K.~Menon laboratory (CREST) for helping us with assembly and storage of payload components in the cleanroom environment.

\bibliographystyle{spiejour}
\bibliography{report}   

\begin{thebibliography}{10}

\bibitem{star}
C.~Liebe, ``Star trackers for attitude determination,'' {\em Aerospace and
  Electronic Systems Magazine, IEEE} {\bf 10}, 10 -- 16  (1995).

\bibitem{rad_level}
T.~Reid, {\em Orbital Diversity for Global Navigation Satellite Systems}.
\newblock PhD thesis  (2017).

\bibitem{Lee20}
S.~Lee, R.~Saleem, and S.-S. Lee, ``Micro star tracker with a curved vane for a
  short baffle length and sharp stray light attenuation,'' {\em Appl. Opt.}
  {\bf 59}, 4131--4142  (2020).

\bibitem{mico}
M.~Sarpotdar, J.~Mathew, A.~Sreejith, {\em et~al.}, ``A software package for
  evaluating the performance of a star sensor operation,'' {\em Experimental
  Astronomy} {\bf 43}  (2016).

\bibitem{mico_cam}
M.~Sarpotdar, J.~Mathew, A.~Sreejith, {\em et~al.}, ``Design and development of
  a star sensor cum asteroid tracker,'' {\em Space Telescopes and
  Instrumentation 2016: Optical, Infrared, and Millimeter Wave, SPIE
  Astronomical Telescopes + Instrumentation} , 9904--195  (2016).

\bibitem{sost}
S.~Gutiérrez, C.~Fuentes, and M.~Diaz, ``{Introducing SOST: An Ultra-Low-Cost
  Star Tracker Concept Based on a Raspberry Pi and Open-Source Astronomy
  Software},'' {\em IEEE Access} {\bf 8}, 166320--166334  (2020).

\bibitem{rasp}
{Raspberry Pi Zero}, ``{Raspberry Pi Foundation.},'' {\em
  http://raspberrypi.org}   (2000).

\bibitem{st-16}
T.~Dzamba, J.~Enright, D.~Sinclair, {\em et~al.}, ``Success by 1000
  improvements: Flight qualification of the st-16 star tracker,''  (2014).

\bibitem{comptech}
B.~Seng, ``{Space Sensor Commercialization – A small company approach}.''
  \url{https://www.osti.gov/servlets/purl/1249037}.

\bibitem{PROCYON&RIGEL}
``{Highly accurate, flexible, robust and scalable multicamera system for
  spacecraft autonomous attitude determination through low cost cameras}.''
  \url{https://ec.europa.eu/research/participants/documents/downloadPublic?documentIds=080166e5adea83b5&appId=PPGMS}.

\bibitem{gv}
M.~Kolomenkin, S.~Pollak, I.~Shimshoni, {\em et~al.}, ``Geometric voting
  algorithm for star trackers,'' {\em Aerospace and Electronic Systems, IEEE
  Transactions on} {\bf 44}, 441 -- 456  (2008).

\bibitem{quest}
G.~Wahba, ``A least squares estimate of satellite attitude,'' {\em Society for
  Industrial and Applied Mathematics} , Problem 65--1  (1966).

\bibitem{hip}
M.~A.~C. Perryman, L.~Lindegren, J.~Kovalevsky, {\em et~al.}, ``The hipparcos
  catalogue.,'' {\em Astronomy and Astrophysics} {\bf 500}, 501--504  (1997).

\bibitem{rpi_cam}
M.~Pagnutti, R.~Ryan, G.~Cazenavette, {\em et~al.}, ``Laying the foundation to
  use raspberry pi 3 v2 camera module imagery for scientific and engineering
  purposes,'' {\em Journal of Electronic Imaging} {\bf 26}, 013014  (2017).

\bibitem{baffle}
M.~Asadnezhad, A.~Eslamimajd, and H.~Hajghassem, ``Optical system design of
  star sensor and stray light analysis,'' {\em Journal of the European Optical
  Society} {\bf 14}  (2018).

\bibitem{api}
R.~M. Salinas, ``{RaspiCam: C++ API for using Raspberry camera with/without
  OpenCv}.'' \url{http://www.uco.es/investiga/grupos/ava/node/40}.

\bibitem{algorithms}
B.~Spratling and D.~Mortari, ``A survey on star identification algorithms,''
  {\em Algorithms} {\bf 2}  (2009).

\bibitem{1976}
P.~SALOMON and W.~GOSS, ``A microprocessor-controlled ccd star tracker,'' {\em
  14th Aerospace Sciences Meeting}   (1976).

\bibitem{tri1}
B.~Quine and H.~F. Durrant-Whyte, ``{Rapid star-pattern identification},'' {\em
  Acquisition, Tracking, and Pointing X, Proc. SPIE} {\bf 2739}, 351 -- 360
  (1996).

\bibitem{tri2}
E.~Heide, M.~Kruijff, S.~Douma, {\em et~al.}, ``Development and validation of a
  fast and reliable star sensor algorithm with reduced data base,'' {\em
  International Astronautical Congress}  (1998).

\bibitem{tri3}
A.~Nabi, Z.~Ahmed-Foitih, and M.~E.-A. Cheriet, ``Improved triangular-based
  star pattern recognition algorithm for low-cost star trackers,'' {\em Journal
  of King Saud University - Computer and Information Sciences} {\bf 33}(3),
  258--267  (2021).

\bibitem{tri4}
G.~Lamy Au~Rousseau, J.~Bostel, and B.~Mazari, ``Star recognition algorithm for
  aps star tracker: oriented triangles,'' {\em IEEE Aerospace and Electronic
  Systems Magazine} {\bf 20}(2), 27--31  (2005).

\bibitem{mortari-py}
D.~Mortari, M.~Samaan, C.~Bruccoleri, {\em et~al.}, ``The pyramid star
  identification technique,'' {\em NAVIGATION} {\bf 51}  (2004).

\bibitem{tri5}
Q.~Fan and X.~Zhong, ``A triangle voting algorithm based on double feature
  constraints for star sensors,'' {\em Advances in Space Research} {\bf 61}(4),
  1132--1142  (2018).

\bibitem{accu-per}
C.~Liebe, ``Accuracy performance of star trackers - a tutorial,'' {\em IEEE
  Transactions on Aerospace and Electronic Systems} {\bf 38}(2), 587--599
  (2002).

\bibitem{triad}
H.~D. BLACK, ``A passive system for determining the attitude of a satellite,''
  {\em AIAA Journal} {\bf 2}(7), 1350--1351  (1964).

\bibitem{q-method}
J.~E. Keat, ``Analysis of least-squares attitude determination routine doaop,''
  {\em Computer Sciences Corporation}   (1977).

\bibitem{svd}
F.~Markley, ``Attitude determination using vector observations and the singular
  value decomposition,'' {\em The Journal of the Astronautical Sciences} {\bf
  36}(3), 245--258  (1988).

\bibitem{thesis}
A.~O. Erlank, {\em Development of CubeStar A CubeSat-Compatible Star Tracker}.
\newblock PhD thesis  (2013).

\bibitem{wahba}
G.~Wahba, ``A least squares estimate of satellite attitude,'' {\em SIAM Review}
  {\bf 7}(3), 409--409  (1965).

\bibitem{1008988}
C.~C. {Liebe}, ``Accuracy performance of star trackers - a tutorial,'' {\em
  IEEE Transactions on Aerospace and Electronic Systems} {\bf 38}(2), 587--599
  (2002).

\bibitem{astrometry}
D.~Lang, D.~W. Hogg, K.~Mierle, {\em et~al.}, ``Astrometry. net: Blind
  astrometric calibration of arbitrary astronomical images,'' {\em The
  Astronomical Journal} {\bf 139}(5), 1782  (2010).

\bibitem{sip}
D.~Shupe, M.~Moshir, J.~Li, {\em et~al.}, ``The sip convention for representing
  distortion in fits image headers,'' {\em ASP Conf. Ser.} {\bf 347}, 491
  (2005).

\bibitem{markley2014fundamentals}
F.~Markley and J.~Crassidis, {\em Fundamentals of Spacecraft Attitude
  Determination and Control}, Space Technology Library, Springer New York
  (2014).

\bibitem{bortle}
J.~E. Bortle, ``Gauging light pollution: The bortle dark-sky scale,'' {\em Sky
  \& Telescope} {\bf Retrieved 2020-05-29}  (2001).

\bibitem{Mathew_2017}
J.~Mathew, A.~Prakash, M.~Sarpotdar, {\em et~al.}, ``{Prospect for UV
  observations from the Moon. II. Instrumental design of an ultraviolet imager
  LUCI},'' {\em Astrophysics and Space Science} {\bf 362}, 1--11  (2017).

\bibitem{Kumar_UVIT_2012}
A.~{Kumar}, S.~K. {Ghosh}, J.~{Hutchings}, {\em et~al.}, ``{Ultra Violet
  Imaging Telescope (UVIT) on ASTROSAT},'' in {\em Space Telescopes and
  Instrumentation 2012: Ultraviolet to Gamma Ray},  T.~{Takahashi}, S.~S.
  {Murray}, and J.-W.~A. {den Herder}, Eds., {\em Society of Photo-Optical
  Instrumentation Engineers (SPIE) Conference Series} {\bf 8443}, 84431N
  (2012).

\bibitem{thermovac}
A.~Kandala, P.~Hari, H.~Simha, {\em et~al.}, ``Design and development of a
  ps4-op payload for solar spectral irradiance measurements and technology
  demonstration of small-satellite subsystems,''  (2021).

\end{thebibliography}
\end{spacing}
\end{document}